\documentclass[conference]{IEEEtran}
\usepackage{multirow}
\usepackage{mathtools}
\usepackage{nicematrix}
\usepackage{float}
\usepackage{tabularx}
\usepackage{booktabs}
\usepackage{cancel}
\usepackage{enumitem} % to use (a) instead of 1)
\usepackage{ytableau}
\usepackage{graphicx} %package to manage images
\graphicspath{ {images/} }

\usepackage{color, colortbl}
\usepackage{multirow}

\usepackage{cuted}
\usepackage[linesnumbered,ruled,vlined]{algorithm2e} % ,noend %remove "vlined" for puting "end"

\usepackage[utf8]{inputenc}
\usepackage[english]{babel}
\usepackage{amsthm,amssymb}
\usepackage{amsmath}
\usepackage{bm}
\usepackage{optidef}
\usepackage{mathrsfs}

\usepackage[font=small,labelfont=bf,labelsep=space]{caption}
\usepackage{xcolor}
\def\BibTeX{{\rm B\kern-.05em{\sc i\kern-.025em b}\kern-.08em
    T\kern-.1667em\lower.7ex\hbox{E}\kern-.125emX}}

\usepackage[style=ieee,backend=biber]{biblatex}
\AtBeginBibliography{\footnotesize} %scriptsize

\usepackage{tikz}
\usetikzlibrary{matrix,backgrounds,positioning}

\usetikzlibrary{calc,patterns,decorations.pathreplacing,decorations.pathmorphing,matrix,positioning,arrows.meta,fit,bending}
\theoremstyle{definition}
\newtheorem{example}{Example}
\newtheorem{theorem}{Theorem}
\newtheorem{proposition}{Proposition}
\newtheorem{lemma}{Lemma}
\newtheorem{corollary}{Corollary}

\newtheorem{definition}{Definition}

\newtheorem{remark}{Remark}

\DeclareMathOperator{\bin}{bin}
\DeclareMathOperator{\supp}{supp}

\DeclareMathOperator{\w}{w}
\DeclareMathOperator{\dist}{d}

\newcommand{\midsepremove}{\aboverulesep = -0.1mm \belowrulesep = -0.1mm}

\newcommand{\wm}{\w_{\min}}

\newcommand{\I}{\mathcal{I}}

\newcommand{\J}{\mathcal{J}}

\newcommand{\C}{\mathcal{C}}

\newcommand{\R}{\mathcal{R}}

\newcommand{\bB}{\mathbf{B}}

\newcommand{\bve}{\bm{\varepsilon}}

\newcommand{\bc}{\boldsymbol{c}}

\newcommand{\bx}{\boldsymbol{x}}

\newcommand{\bv}{\boldsymbol{v}}

\newcommand{\bG}{\boldsymbol{G}}

\newcommand{\ind}{\operatorname{ind}}
\newcommand{\ev}{\operatorname{ev}}

\newcommand{\ff}{\mathbb{F}}
\newcommand{\ft}{\mathbb{F}_2}

\definecolor{c1}{RGB}{250, 250, 250}
\definecolor{c2}{RGB}{250, 250, 200}
\definecolor{c3}{RGB}{250, 250, 150}
\definecolor{c4}{RGB}{250, 250, 100}
\definecolor{c5}{RGB}{250, 250, 50}
\definecolor{c6}{RGB}{250, 250, 1}
\definecolor{c7}{RGB}{250, 200, 1}
\definecolor{c8}{RGB}{250, 150, 1}
\definecolor{c9}{RGB}{250, 100, 1}
\definecolor{c10}{RGB}{250, 50, 1}
\definecolor{c11}{RGB}{250, 1, 1}
\definecolor{c12}{RGB}{200, 1, 1}
\definecolor{c13}{RGB}{150, 1, 1}
\definecolor{c14}{RGB}{100, 1, 1}
\definecolor{c15}{RGB}{50, 1, 1}

\newcommand{\Alow}{{\rm LTA}(m,2)}

\newcommand{\GL}{{\rm GL}(m,2)}
\newcommand{\Mon}{\mathcal{M}_{m}}
\newcommand{\weako}{\preceq_w}
\newcommand{\Rm}{{\mathbf {R}}_m}
\newcommand{\Aut}{{\mathscr {Aut}}}

\IEEEoverridecommandlockouts                              % This command is only
\title{%Partial
Weight Structure of Low/High-Rate Polar Codes and Its Applications
}

\author{
\IEEEauthorblockN{Mohammad Rowshan$^{\ast\dagger}$, Vlad-Florin Drăgoi$^{\ast\ddagger}$, and Jinhong Yuan$^\dagger$, {\em Fellow, IEEE}}
\IEEEauthorblockA{$^\dagger$School of Electrical Eng. and Telecom., University of New South Wales (UNSW), Sydney, Australia\\
$^\ddagger$Faculty of Exact Sciences, Aurel Vlaicu University, Arad, Romania\\
\{m.rowshan,j.yuan\}@unsw.edu.au, vlad.dragoi@uav.ro }
\thanks{$^\ast$These authors contributed equally (Corresponding author: Vlad-Florin Drăgoi).}%, vlad.dragoi@uav.ro).}
}

\begin{document}

\maketitle
\pagestyle{plain}
%\pagestyle{empty}

%%%%%%%%%%%%%%%%%%%%%%%%%%%%%%%%%%%%%%%%%%%%%%%%%%%%%%%%%%%%%%%%%%%%%%%%%%%%%%%%
\begin{abstract}
The structure of a linear block code is pivotal in defining fundamental properties, particularly weight distribution, and code design. In this study, we characterize the Type II structure of polar codewords with weights less than twice the minimum weight $\wm$, utilizing the lower triangular affine (LTA) transform. We present a closed-form formula for their enumeration. Leveraging this structure and additionally characterizing the structure of weight $2\wm$, we ascertain the complete weight distribution of low-rate and, through the utilization of dual codes properties, high-rate polar codes, subcodes of Reed--Muller (RM) codes, and RMxPolar codes.  Furthermore, we introduce a partial order based on the weight distribution and briefly explore its properties and applications in code construction and analysis. 
% \vlad{1.Partial order wrt. Weight distribution\\
% 2. Algorithm for RM-Polar based on Weight contribution\\
% 3. Type II codewords for small dimension RM-Polar codes\\
% 4. What weights do there exist for such codes?\\
% 5. The complete weight distribution for small dimension RM-Polar.
% 6. Compare with CRC-Polar (the 6bits CRC for dimension 12, 19), the RM-Polar generated based on weight contribution, and the CRC-RM-Polar
% 7. The impact of the reduction in the minimum weight codewords on other weights -- design}

% {\color{red}Contributions:\\
% 1. We characterize Type II codewords \\
% 2. Give closed-form formula for Type II\\
% 3. All weight distribution for the subcode of RM (2,m)\\
% 4. Propose a finner RM-Polar construction (maybe merge it into item 6)\\
% 5. Discuss duality and low to high rate properties. Does McWilliams and SLoane formulae induce transfer of best code wrt. reliability and weight distribution.
% 6. POSET of weight contributions and its relation with POSET of reliability.}
\end{abstract}


%\cite{arikan}

\section{Introduction}
Polar codes, identified as a code family capable of achieving the capacity of any binary-input discrete memoryless channel (BI-DMC) \cite{arikan}, are members of a larger family of codes called decreasing monomial codes \cite{bardet2016crypt}, which also includes Reed--Muller codes. While significant research has been devoted to polar codes in the past decade, mainly focused on reducing the decoding complexity or improving the error correction performance, less attention has been paid to their structural aspects. Building on classical results pertaining to Reed-Muller codes \cite{kasami1970weight,kasami1976w2.5d}, recent studies \cite{bardet} and \cite{rowshan1.5w} have delved into the algebraic properties of polar codes and, in general, decreasing monomial codes, among which a subgroup of their permutation group called the lower triangular affine group ($\Alow$). These works shed light on the weight structure, offering closed-form formulas for the enumeration of codewords with a weight less than twice the minimum weight ($\wm$). The approach used in these works involves the action of $\Alow$ on the maximum-degree monomials in the construction set and the cardinality of the resulting orbit of polynomials, which include $\wm$-weight codewords. Furthermore, they provide insight into the $1.5\wm$-weight codewords through the Minkowski sum of two particular orbits. The permutation group and the discovery of its properties \cite{li2021complete, PBL22, IU22,IU22a, pillet2023distribution} have made a significant contribution to the ensemble decoding of polar codes \cite{GEECB21, GEEB21}. 

In this study, we extend the findings presented in \cite{rowshan1.5w} to characterise larger weights. Due to space constraints, our focus centers on the Type II structure \cite[Thm. 2]{rowshan1.5w}, for which we provide a closed-form formula to enumerate its codewords. We demonstrate that this structure yields nearly the entire weight distribution of the second-order Reed-Muller codes $\R(2,m)$ and their subcodes. We also propose a formula for the codewords of weight $2\wm$ to complete the weight distribution.  Additionally, we leverage this structure and the associated closed-form formula to construct low-rate RM-polar codes \cite{rowshan-pac1} based on $\R(2,m)$ subcodes. We extend this construction to high-rate codes by exploiting the properties of the corresponding dual codes. Finally, we introduce a new partial order based on weight distribution, analyse its properties, and give examples for its potential application for code construction. %The complete manuscript with full proofs will be available on arXiv.
%%%%%%%%%%%%%%%%%%%%%%%%%%%%%%%%%%%%%%%%%%%%%%%%%%%%
\section{Polar codes as Decreasing monomial codes}%\mohammad{We need to make this section as short as possible and if needed by referring the readers to other references. We decide on the cuts when we complete the other sections.}
\subsection{Coding theory notations}
We denote by $\ft$ the finite field with two elements. 
Also, subsets of consecutive integers are denoted by $[\ell,u]\triangleq\{\ell,\ell+1,\ldots,u\}$. %, we use the notation $[n]\triangleq[0,1,\dots,n-1]$. %The binary representation of an integer $i \in [0,2^n-1]$ is defined as  $\bin(i)=\mathbf{i}=i_{n-1}...i_1i_0$, where $i_0$ is the least significant bit, that is $i = \sum_{a=0}^{n-1}i_a 2^a$. 
%We define below standard notions from coding theory (for instance, see \cite{lin_costello}).
The \emph{support} of a vector $\bc = [c_0,\ldots,c_{N-1}] \in \ft^N$ is definend as $\supp(\bc) \triangleq \{i \in [0,N-1] \mid c_i \neq 0\}$. %\mohammad{I can change $\supp$ to $\ind$ in the last section because we have used $\ind$ more throughout the paper.} 
%\vlad{We have used $\ind$ for monomials and $\supp$ for vectors.} %\mohammad{Yes, but I noticed that you commented the definition of $\ind$ for monomials in the following.} %The support for monomials where every variable represents a zero coordinate, $c_i=0$, is denoted by $\ind(\bc)$. %Let $\Si$ denote the support of $\bin(i)$, that is, $\Si\triangleq\supp(\bin(i))$. 
The cardinality of a set is denoted by $|\cdot|$ and the set difference by $\backslash$. %For instance, $\Sj\backslash\Si$ gives all the elements in the support of $\bin(j)$ that are not in the support of $\bin(i)$. 
The Hamming \emph{weight} of a vector $\bc \in \ft^N$ is $\w(\bc)\triangleq |\supp(\bc)|$. %or $w(\mathbf{c})=d(\mathbf{c},0)$. 
% %
The binary vectors $\bin(i)=(i_0,\dots,i_{m-1})\in \ft^m$, where $i=\sum_{j=0}^{m-1} i_j2^j$, are written so that $i_{m-1}$ is the most significant bit. In addition, we endow $\ft^m$ with the decreasing index order relation (see Defs. \ref{def:order},\ref{def:dec_set}).
%To any $\bi\in \ft^m$ we associate its integer value $i=\sum_{j=0}^{m-1} i_j2^j$ (the sum is computed over $\mathbf{N}$) and we order the integers in $[0,2^m-1]$ w.r.t. the decreasing index order. 
For example, the elements in $\ft^2$ are $\{(1,1),(0,1),(1,0),(0,0)\}.$ 

A $K$-dimensional subspace $\C$ of $\ft^N$ is called a linear $(N,K,d)$ \emph{code} over $\ft$ 
where $d$ is the minimum distance of code $\C$; that is,
\begin{equation*}
\dist(\C) \triangleq \min_{\bc,\bc' \in \C, \bc \neq \bc'} \dist(\bc,\bc')=\wm,
\end{equation*}
 which is equal to minimum weight of codewords excluding all-zero codeword.
% It is easy to see that the Hamming norm induces the Hamming distance and vice-versa. Hence, we have (see \cite[Section 3.3]{lin_costello})
%  \[\wm\triangleq\min_{\bc\in\C,\bc\neq 0}\w(\bc)=\dist(\C).\] 
We usually use the short notation $(N,K)$ for codes where we refer to $N$ and $K$ as the \emph{length} and the \emph{dimension} of the code. The vectors in $\C$ are called \emph{codewords}.  
Let us collect all $\w$-weight codewords of $\C$ in set $W_{\w}$ as
$
    W_{\w}(\C) = \{ \bc\in\C \mid \w(\bc)=\w \}.
$
A \emph{generator matrix} $\bG$ of an $(N,K)$-code $\C$ is a $K \times N$ matrix in $\ft^{K\times N}$ whose rows are $\ft$-linearly independent codewords of $\C$. Then $$\C = \{\bv \bG \colon \bv \in \ft^K\}.$$

\paragraph{Multivariate monomial and polynomials}
Let $m$ be a fixed integer that represents the number of different variables $\mathbf{x}\triangleq(x_{0},\dots,x_{m-1})$ and $\ft[x_0,\dots{},x_{m-1}]$ be the set of polynomials in $m$ variables. %Since we are dealing with binary codes, we will identify $x_i$ with $x_i^2$ (using the Frobenius endomorphism) and 
Also, consider the ring $\Rm=\ft[x_0,\dots{},x_{m-1}]/(x_0^2-x_0,\dots,x_{m-1}^2-x_{m-1})$. 
A particular basis for $\Rm$ is the monomial basis, where any $m$-variate monomial can be defined as $\mathbf{x}^{\bin(i)}=x_0^{i_0}\cdots x_{m-1}^{i_{m-1}},$
%$$\mathbf{x}^{\bin(i)}=\prod_{j=0}^{m-1}x_{j}^{i_{j}}=x_{0}^{i_{0}} \cdots x_{m-1}^{i_{m-1}},$$ 
where $\bin(i)=(i_{0},\dots,i_{m-1})$ with $i_j\in\{0,1\}.$ 
Denote the set of all monomials in $\Rm$ by 
$
\Mon\!\triangleq\!\left\{\mathbf{x}^{\bin(i)} \mid\bin(i) \in \mathbb{F}_{2}^{m}\right\}
.$
One can associate to any integer its binary expansion, which is a one-to-one mapping, and thus map each integer $i\in[0,2^m-1]$ into a monomial $\mathbf{x}^{{\bin(i)}}.$ 
 Given that the reverse index order was used for $\ft^m$, we will consider the bijection:  
% \[
% i\in[0,2^m-1] \mapsto \mathbf{x}^{{\bin(2^m-1-i)}}\in\Mon.
% \]
\[
\begin{array}[h]{ccccc}
[0,2^m-1]    & \to &\Mon\\
i& \mapsto &\mathbf{x}^{{\bin(2^m-1-i)}}
\end{array}
\]
For example, for $m=2$, we have $\Mon=\{x_0x_1,x_1,x_0,\bm{1}\}.$ 

Furthermore, the \emph{support of monomial} $f=x_{l_1}\dots x_{l_s}$ is $\ind(f)=\{l_1,\dots,l_s\}$ and its degree is $\deg(f)=|\ind(f)|.$ The degree induces a ranking on any monomial set $\I\subseteq\Mon$, that is, $\I=\bigcup_{j=0}^{m}\I_j$, where $\I_j=\{f\in\I\mid\deg(f)=j\}.$

\paragraph{Monomial codes}
Let us define the evaluation function denoted by 
$\ev(g)$ for a polynomial $g \in \Rm$ that associates $g$ with a binary vector in $\ft^{2^m}.$ 
\begin{equation}
\begin{array}[h]{ccccc}
%\Rm    & \to &\ft^{2^m}\\
g& \mapsto &\ev(g) = \big(g(\bin(i)) \big)_{\bin(i) \in \ft^m}
\end{array}
\end{equation}
% \begin{definition}
% Let $g\in \Rm$ and arrange the elements in $\ft^m$ in decreasing index order. We define the evaluation function:
% \[
% \begin{array}[h]{ccccc}
% %\Rm    & \to &\ft^{2^m}\\
% g& \mapsto &\ev(g) = \big(g(\bin(i)) \big)_{\bin(i) \in \ft^m}
% \end{array}
% \]
% \end{definition}

The function $\ev(\cdot)$ defines a vector space isomorphism between the vector space $(\Rm,+,\cdot)$ and $(\ft^{2^m},+,\cdot).$

\begin{example}
For $m=3$ and $g=x_0x_1+x_0$ we have 
\begin{center}{\scriptsize
\begin{tabular}{lccccccccc}
&&111&011&101&001&110&010&100&000\\
\cline{3-10}
$\ev(x_0x_1)$&=&1&0&0&0&1&0&0&0\\
$\ev(x_0)$&=&1&0&1&0&1&0&1&0\\
\hline
$\ev(g)$&=&0&0&1&0&0&0&1&0\\
\end{tabular}
}
\end{center}
\end{example}

%We can now define monomial/polynomial codes. 
\begin{definition}[Monomial code]
Let $\I\subseteq\Mon.$ be a finite set of polynomials in $m$ variables. 
 A monomial code defined by $\I$ is the vector subspace $\C(\I) \subseteq \ft^{2^m}$ 
generated by $\{ \ev(f) ~|~ f \in \I\}$.

% \begin{itemize}
%  \item When $\I\subseteq \Rm$ we say that $\C(\I)$ is a \emph{polynomial code}.

% \item When $\I\subseteq \Mon$ we say that $\C(\I)$ is a \emph{monomial code}.
% \end{itemize}
\end{definition}

%To demonstrate onstructing monomial codes, we
Let $\bG_{2^m} \triangleq \begin{pmatrix} 1 & 0 \\ 1 & 1 \end{pmatrix}^{\otimes m}$, where $\otimes m$ denotes the Kronecker power. 
% \[
% \bG_{2^m} \triangleq \underbrace{\begin{pmatrix} 1 & 0 \\ 1 & 1 \end{pmatrix} \otimes \cdots 
% \otimes \begin{pmatrix} 1 & 0 \\ 1 & 1 \end{pmatrix}}_{m \;\text{ times}}.
% \]
In \cite{dragoi17thesis}, it was shown that $\bG_{2^m}$ is a basis for the vector space $\ft^{2^m}$, based on the mapping% This fact is grounded in the~following:
\[
\begin{array}[h]{ccccc}
[0,2^m-1]    & \to &\Mon   & \to &\ft^{2^m}\\
i& \mapsto &g\triangleq\mathbf{x}^{{\bin(2^m-1-i)}} & \mapsto &\bG_{2^m}[i]=\ev(g) 
\end{array}
\]
Table \ref{tab:G8_eval} illustrates the evaluation of all rows of matrix $\bG_8$ as an example. 
\begin{table*}[ht]
    \centering{
\begin{tabular}{cccc||c c c c c c c c}
Index - $i$&$\bin(i)$&$\bin(2^m-1-i)$&$g$ & 111&011&101&001&
110& 010& 100& 000\\  
\hline
\hline
0&(000)&(111)&$\ev(x_0x_1x_2)$&1&0&0&0& 0&0&0&0\\
1&(100)&(011)&$\ev(x_1x_2)$&1&1&0&0&0&0&0&0\\
2&(010)&(101)&$\ev(x_0x_2)$&1&0&1&0&0&0&0&0\\
3&(110)&(001)&$\ev(x_2)$&1&1&1&1& 0&0&0&0\\
4&(001)&(110)&$\ev(x_0x_1)$ &1&0&0&0&1&0&0&0\\
5&(101)&(010)&$\ev(x_1)$&1&1&0&0&1&1&0&0\\
6&(011)&(100)&$\ev(x_0)$&1&0&1&0&1&0&1&0\\
7&(111)&(000)&$\ev(1)$&1&1&1&1& 1&1&1&1\\
\end{tabular}
}
    \caption{The matrix $\bG_8$ as evaluation of monomials in $\Mon$. %\mohammad{In polar coding literature including Ar\i kan's paper, we have $\bG_{2^m}$.}
    }
    \label{tab:G8_eval}
\end{table*}

\paragraph{Decreasing monomial codes}\label{par:dec_monomial_codes}

Let ``$|$" denote the divisibility between monomials, i.e., $f|g$ iff $\ind(f)\subseteq\ind(g).$ Also, the greatest common divisor of two monomials is $\gcd(f,g)=h$ with $\ind(h)=\ind(f)\cap\ind(g).$

\begin{definition}\label{def:order}Let $m$ be a positive integer and $f,g\in\Mon.$ Then $f\weako g$ if and only if $ f|g.$ When $\deg(f)=\deg(g)=s$ we say that $f\preceq_{sh} g$ if $\forall\;1\leq\ell\leq s\;\text{ we have }\;  i_\ell \le j_\ell$, where $f=x_{i_1}\dots x_{i_s}$, $g=x_{j_1}\dots x_{j_s}$. 
Define $f\preceq g\quad \text{iff}\quad \exists g^*\in \Mon\;\text{s.t.}\; f\preceq_{sh} g^*\weako g$.
\end{definition}

Note that we have $x_0\preceq x_1\preceq \dots \preceq x_{m-1}.$ Also, $\preceq$ is an order relation that is partial, e.g., $x_3x_4$ and $x_1x_5$ are not comparable with respect to $\preceq.$ 
Now we can define the concept of monomial decreasing sets.

\begin{definition}\label{def:dec_set}
      A set $\I \subseteq \Mon$ is \emph{decreasing}  if and only if ($f \in \I$ and $g \preceq f $) implies $g \in \I$. 

%      A decreasing closed interval with respect to $\preceq$ is $[f,g]_{\preceq}=\{h\in \Mon \mid  f\preceq h\preceq g\}.$%\vlad{I think we have removed all the intervals! If true, we do not need it any more.}
\end{definition}

Any monomial code $\C(\I)$ with $\I$ decreasing is called a \emph{decreasing monomial code}. Both Polar and Reed-Muller codes are decreasing monomial codes \cite{bardet}. For $\R(r,m)$ we have
\begin{equation}
\label{eq:RM_decreasing}
 \R(r,m)=\C\left([1, x_{m-r}\cdots{} x_{m-1}]_{\preceq}\right).%=\CC(\{x_{m-r}\cdots{} x_{m-1}\})_{\preceq}.
\end{equation}

From \cite{bardet}, we use the multiplicative complement of a monomial $g\in \Mon$ defined by $\check{g}\triangleq \frac{x_0\dots x_{m-1}}{g}.$ Note that while $g=x^{\bin(2^m-1-i)}$ for a given $i$, its complement is $(\check{g})=x^{\bin(i)}.$ We can extend the definition to a monomial set $\check{\I}=\{\check{g}\mid g\in \I\}.$ The dual of a decreasing monomial code is still a decreasing monomial code: 
$\C(\I)=\C(\Mon\setminus \check{\I})$.

\subsection{Permutation group}\label{ssec:perm_grp}
The set of permutations that globaly leave the code invariant, forms the \emph{automorphism group} of the code $\C.$ 
A bijective affine transformation over $\ft^m$ is represented by a pair $(\bB, \bm{\varepsilon})$ where $\bB=(b_{i,j})\in \ff_2^{m\times m}$ is an invertible matrix lying in the general linear group $\GL$ and $\bve$ in $\ft^m$. The action of $(\bB, \bve)$ on a monomial $g=\prod_{i\in\ind(g)}x_i$ (denoted by $(\bB, \bve)\cdot g$) replaces each variable $x_i$ of $g$ by a variable $y_i$ as
$y_i = x_i + \sum_{j=0}^{i-1} b_{i,j} x_j + \varepsilon_i$. 
%y_i = x_i + \sum_{j=0,j\not\in\ind(\bg)}^{i-1} a_{i,j} x_j + \epsilon_i,
% $
% where $b_{i,j}$ and $\varepsilon_i$ are in $\ft$. 
This new variable $y_i$, is in fact a linear form (a polynomial in which all terms have a degree at most 1), where the maximum variable is $x_{i}.$%, w.r.t. the order relation $\preceq.$

In \cite{bardet} it was proved that the lower triangular affine transformation denoted by $\Alow$ is a subgroup of the permutation group of any decreasing monomial code. $\Alow$ is a subgroup of the affine group, where $\bB\in\GL$ is a lower triangular binary matrix with $b_{i,i}=1$ and $b_{i,j}=0$ whenever $j>i$. Therefore, the action of $\Alow$ can be expressed as the following mapping from $\ft^m$ to itself: 
$
\bx \rightarrow \bB \bx + \bve.
$
The set of polynomials resulting from the action of $\Alow$ on a monomial is collected in a set named \emph{orbit} and denoted by
$\Alow \cdot f = \{(\bB,\bve) \cdot f\mid (\bB,\bve) \in \Alow\}.$
    
Since $\Alow$ acts as a permutation on $\ev(f)$, all elements in $\Alow\cdot f$ have the same Hamming weight. %This 

\subsection{Minimum Weight Codewords}\label{ssec:wmin_orbit}

The minimum weight codewords of any monomial code $\C(\I)$ can be written as evaluations of particular polynomials. Indeed, letting $r=\max\{\deg(f)\mid f\in \I\}$ we have that $\wm=2^{m-r}$ and any $\bc\in\C(\I)$ of weight $\wm$ equals $\ev(l_1\cdots l_r)$, where $l_i$ are linear forms mutually independent. In other words one can not write $l_i$ as a linear combination of the other forms $l_j$ (for $i\neq j$), and each $\deg(l_i)=1$ for all $i.$ Notice that any such linear form has a maximum variable w.r.t. $\preceq$, i.e., we can always write $l_i=x_{i_1}+\dots x_{i_l}+\epsilon_i$ where $x_{i_l}\preceq \dots x_{i_1}$ and $\epsilon_i\in \ft.$

Minimum weight codewords are counted using a particular subgroup of $\Alow$.

\begin{definition}[\cite{bardet,dragoi17thesis}] For any $g\in\Mon$ define $\Alow_g$ as the subgroup $\Alow$ by "collecting actions
 of $(\bB,\bve)$ that satisfy" %\mohammad{I am a bit confused with various notations. I thought we wanted to change all notations $\Ab$ to $\Be$ as used in Example 9, Proposition 1, etc. Should I proceed with the changes?}
 %\vlad{I have changed to $\bA,\bve$. I will do the modifications.}
 \vspace{-5pt}
\[
 \varepsilon_i = 0 \text{ if } i \not \in \ind(g) 
 ~~~\text{ and }~~ 
 b_{ij} = 
 \left \{ 
 	\begin{array}{l}
		0 \text{ if } i \not \in  \ind(g), \\
		0 \text{ if }  j \in \ind(g).
	 \end{array}
 \right.
\] 
\end{definition}

% \begin{theorem}[\cite{bardet,dragoi17thesis}]\label{thm:min-weight}
% Let $f\in \Mon.$ Then we have \begin{equation}
% \Alow\cdot f=\Alow_f\cdot f.
% \end{equation}
% \end{theorem}
In \cite{bardet} it was demonstrated that $\Alow\cdot f=\Alow_f\cdot f$, hence, $\Alow_f\cdot f$ being a finer estimation of the orbit. This group can be written as a semi-direct product between the group of invertible matrices having one on the diagonal and the group of translations. A variable $x_i$ can be translated by a scalar $\varepsilon_i\in\ft$ onto $x_i+\varepsilon_i$. Hence, $f\in\Mon$ admits $2^{\deg(f)}$ translations. 
Regarding the linear mapping, $x_i$ can be transformed into a "new variable" ($y_i$), where $y_i=x_i+\sum_{j=0, j \notin \operatorname{ind}(f)}^{i-1} b_{i,j} x_{j}$. The extra variables considered in $y_i$ express the degree of freedom we have on $x_i$ and are denoted by $\lambda_f(x_i)=|\{j\in[0,i) \mid j\notin\ind(f)\}|.$ 
\begin{definition}\label{def:lambda_f}
    Given a monomial $f=x_{i_{0}}\dots x_{i_{s-1}}$ with $i_0<\dots< i_s$, we associate a partition of length $\deg(f)$ defined by $\lambda_f=(i_{s-1}-(s-1),\dots,i_0-0)$ (see \cite{dragoi17thesis}). Then, the total number of free variables on all $x_i$ in $f$ is 
    \begin{equation}\label{eq:lambda_f}        |\lambda_f(f)|=\sum_{i\in\ind(f)}\lambda_f(x_i)
    \end{equation}
    from which 
    $2^{|\lambda_f|}$ possible actions on $f.$ For simplicity, we use the notation $\lambda_f$ for $\lambda_f(f)$. 
\end{definition}
%To each monomial $f=x_{i_{0}}\dots x_{i_{s-1}}$ with $i_0<\dots< i_s$ one can associate a partition of length $\deg(f)$ defined by $\lambda_f=(i_{s-1}-(s-1),\dots,i_0-0)$ (see \cite{dragoi17thesis}). The total number of free variables on all $x_i$ in the support of $f$ is $|\lambda_f(f)|=\sum_{i\in\ind(f)}\lambda_f(x_i)$, (or simply $\lambda_f$) from which     $2^{|\lambda_f|}$ (since we are defined over $\ff_2$) possible actions on $f.$ 
\begin{example}
    Let $f=x_0x_2x_4.$ Then $\lambda_f=(4-2,2-1,0-0)=(2,1,0)$ which corresponds to the degree of freedom 2 for $x_4$ ($x_3,x_1$ are free variables for $x_4$), and the degree of freedom $1$ for $x_2$ ($x_1$ is a free variable for $x_2$). As another example, let $g=x_2x_4.$ Then $\lambda_g=(3,2)$ because $x_3,x_1,x_0$ are free variables for $x_4$ and $x_1,x_0$ are free variables for $x_2.$  We also have $\lambda_f(g)=(2,1)$ since $g|f$ and $\ind(g)=\{2,4\}$ which correspond to the first and second index in $\ind(f).$
\end{example}
Definition \ref{def:lambda_f} can be extended to any monomial $g=x_{j_{0}}\dots x_{j_{l-1}}$ satisfying $g|f$, that is, $\lambda_f(g)$ is the partition of length $l$ defined by $\lambda_f(g)=(\lambda_f(x_i))_{i\in\ind(g)}$, which yields 
\begin{equation}\label{eq:lambda_f_g}
    |\lambda_f(g)|=\sum_{i\in\ind(g)}\lambda_f(x_i).
\end{equation}

Eventually, we get the well known formula from \cite{bardet}% and the enumeration of minimum weight codewords of a de creasing monomial code.
\begin{equation}\label{eq:A_wm}
    \left|\Alow_f \cdot f\right| = 2^{\deg(f)+|\lambda_f|},
\end{equation}
\begin{equation}\label{eq:sum_A_wm}
    |W_{2^{m-r}}(\I)|=\sum\limits_{f\in \I_r}2^{r+|\lambda_f|}.
\end{equation}

%\clearpage
\section{Structure of weights less than $2\wm$}%\mohammad{At this stage people do not know what type II is. Better to choose a more general title.}

The starting point of our quest is a well-known result on the classification of codewords of weight smaller than $2\wm$ of Reed-Muller codes.  
\begin{theorem}
[{\cite{sloane1970weight},\cite[Theorem 1]{kasami1970weight}}]
\label{thm:Kasami-Tokura}
 Let $r<m$ and $P\in\Rm$ be such that $\deg(P)\leq r$ with $0<\w(\ev(P))<2^{m+1-r}.$ Then $P$ is affine equivalent to one of the forms 
\begin{enumerate}
	\item \textbf{Type I}: $P=y_1\dots y_{r-\mu}(y_{r-\mu+1}\dots y_{r}+y_{r+1}\dots y_{r+\mu})$ where $m\geq r+\mu,r\geq \mu\geq 3$
	\item \textbf{Type II}: $P=y_1\dots y_{r-2}(y_{r-1}y_{r}+\dots+y_{r+2\mu-3}y_{r+2\mu-2})$ where $m-r+2\geq 2\mu, 2\mu\geq 2.$
\end{enumerate} 

In both cases $y_i$ are linear independent forms and $\w(\ev(P))=2^{m+1-r}-2^{m+1-r-\mu}.$
\end{theorem}

\subsection{Classification of Type II codewords}
Considering the following fact, we rewrite Theorem \ref{thm:Kasami-Tokura} in terms of $\Alow$ to add more depth to its impact on decreasing monomial codes: Any product of independent linear forms can be rewritten so that the maximum degree variables of these forms are all distinct \cite[Proposition 3.7.12]{dragoi17thesis}. For example, $(x_4+x_2)(x_4+x_1)=(x_4+x_1)(x_2+x_1+1).$ %Leveraging this fact we provide an alternative formulation of Theorem \ref{thm:Kasami-Tokura} in terms of $\Alow.$
\begin{theorem}\label{thm:Kasami-Tokura-1}
    Let $r<m$ and $P\in\Rm$ be such that $\deg(P)\leq r$ with $0<\w(\ev(P))<2^{m+1-r}.$ Then 
\begin{enumerate}
	\item \textbf{Type I:} for $m\geq r+\mu,r\geq \mu\geq 3$\\
    $P=y_1\dots y_{r-\mu}(y_{r-\mu+1}\dots y_{r}+y_{r+1}\dots y_{r+\mu})$\\
    $P\in \Alow\cdot f+\Alow\cdot g$ with $f,g\in \Mon$ having $\deg(f)=\deg(g)=r.$  
	\item \textbf{Type II:} for $m-r+2\geq 2\mu\geq 2$\\
    $P=y_1\dots y_{r-2}(y_{r-1}y_{r}+\dots+y_{r+2\mu-3}y_{r+2\mu-2})$\\
    $P\in \sum_{i=1}^{\mu}\Alow\cdot f_i$ with $f_i\in\Mon$ satisfying $\deg(f_i)=r.$
\end{enumerate} 

In both cases, $y_i$ are linear independent forms and $\w(\ev(P))=2^{m+1-r}-2^{m+1-r-\mu}.$ 
\end{theorem}

In the sequel, we shall focus on Type II codewords. Notice that such codewords are defined by $r+2\mu-2$ independent linear forms. 
The particular case of $\mu=2$, which corresponds to $1.5\wm$ codewords, was determined in \cite{rowshan1.5w}. The authors classified any such codewords in terms of monomials and group action. Let us recall their result. 
\begin{theorem}[\cite{rowshan1.5w}]\label{thm:1.5d}
    Let $\C(\I)$ be a decreasing monomial code, and $r=\max_{f\in \I}\deg(f)$. Any codeword of weight $1.5\wm$ in $\C(\I)$ is the evaluation of a polynomial $P\in\Alow_h\cdot h\cdot\big(\Alow_{f} \cdot \frac{f}{h}+\Alow_{g}\cdot \frac{g}{h}\big)$, with $f,g\in \I_r, h=\gcd(f,g)$, and $\deg(h)=r-2$.
\end{theorem}

Therefore, the $\wm$ and $1.5\wm$ codewords can be described using $\I_r.$ If any such codeword falls within an orbit $\Alow\cdot f$ (for the case of $\wm$) or a sum of two orbits $\Alow\cdot f+\Alow\cdot g$ (for the case of $1.5\wm$) with $f,g\in\I_r$, the question arises whether analogous results exist for higher weight. Continuing along the same path, we prove that any Type II codeword of a decreasing monomial code resides within the sum of $\mu$ orbits of monomials from $\I_r.$ 

\begin{theorem}\label{thm:equality-orbits-typeII}
    Let $\C(\I)$ be a decreasing monomial code and $r=\max_{f\in \I}\deg(f)$. Then, any Type II codeword of weight $2^{m+1-r}-2^{m+1-r-\mu}$ with $m-r+2\geq 2\mu\geq 2$ belongs to 
        \[\Alow_h\cdot h\cdot\sum_{i=1}^{\mu}\Alow_{f_i} \cdot \frac{f_i}{h}\] 
        where $\forall i, f_i\in \I_r$, $h=\gcd(f_i,f_j)$, for all $i,j\in[1,\mu]$ ($i\neq j$) and $\deg(h)=r-2$.
\end{theorem}

Straightforward, given a decreasing monomial code defined by the monomial set $\I$, one can enumerate the possible weights in the interval $[\wm,2\wm].$ Note that for any $\R(r,m)$ code, we have $\mu\leq \lfloor\frac{m-r+2}{2}\rfloor$ given in Theorem \ref{thm:Kasami-Tokura}.

\begin{example}
    $\R(2,7)$ has weights $\{32,48,56,64\}$ which correspond to $\wm=2^{5},1.5\wm=2^{6}-2^{4}, 1.75\wm=2^{6}-2^{8}, 2\wm=2^{6}.$ Considering $m-r+2=7$ for Type II, observe that we can have $\mu\in\{2,3\}$ corresponding to the number of terms in $x_0x_1+x_2x_3$ and $x_0x_1+x_2x_3+x_4x_5.$

    %Let us consider another example with 
    As additional examples, for $m=7,r=3$, let $\I_r=\{x_0x_1x_2,x_0x_1x_3,x_0x_1x_4\}.$ In this case, there are no pair or triplet of monomials, corresponding to $\mu\in\{2,3\}$, with $\deg(\gcd(f_1,f_2))=1$ or $\deg(\gcd(f_1,f_2,f_3))=1$.\\
    Now, let us change the set to $\I_r=\{x_0x_1x_2,x_0x_1x_3,x_0x_1x_4,$ $x_1x_2x_3,x_0x_1x_5,x_1x_2x_4, x_0x_2x_3\}.$ Here, %we have $\mu\leq 2$ because 
    we can find $(x_0x_1x_5,x_0x_2x_3)$ as a valid pair for $\mu=2.$ %\mohammad{I believe it is better not to say "there is no possible value for $\mu$". Instead, we can say that for $mu=2$, we can find a pair with gcd=1. The reason is that one can say the range of mu is defined by the code parameters, bit set $\I$. I am not sure about this but it could be confusing for the readers.}
\end{example}

%\paragraph{Sums of orbits}
\subsection{Enumeration formulae for Type II codewords}

A natural question that arises is the count of polynomials within an orbit, like %that comes into mind relates to how many polynomials one has inside an orbit, such as 
the one from Theorem \ref{thm:equality-orbits-typeII}. It is worth noting that we use the term 'orbit' with a slight abuse of language, since there is a Minkowski sum of multiple orbits. %Since we deal with polynomials of the form $\sum\Alow_{f_i}\frac{f_i}{h}$ where $\deg\left(\frac{f_i}{h}\right)=2$ we first analyse the polynomials inside the sum. Our result generalises the case $\mu=2$ from \cite{rowshan1.5w}. %, more exactly \emph{collisions} of polynomials. In \cite{rowshan1.5w} collisions are defined with respect to addition. Here we will also require collisions for multiplications. 
%The main difficulty resides in estimating the number of polynomials/monomials that can be written as sums of $\mu$ distinct polynomials.  
% \begin{example}
%    Let $f=x_2x_6,g=x_3x_5$ and $P=(x_2+x_1)(x_6+x_4+1)\in \Alow\cdot f$ and $Q=(x_3+x_2+x_0+1)(x_5+x_4+x_2)\in\Alow\cdot g$. We can create two non-trivial distinct pairs $P^{*},Q^{*}$ 
%     \begin{itemize}
%         \item $P^{*}=(x_2+x_1)(x_6+x_4+x_3+x_2+x_0)$ and $Q^{*}=(x_3+x_2+x_0+1)(x_5+x_4+x_1)$ 
%         \item $P^{*}=(x_2+x_1)(x_6+{\color{blue}x_5}+x_3+x_2+x_1+x_0)$ and $Q^{*}=(x_3+x_1+x_0+1)(x_5+x_4+x_1)$
%     \end{itemize}
%  Let $f=x_2x_4,g=x_0x_3.$ We obtain that $f+g=x_2(x_4+x_0)+x_0(x_3+x_2).$  
% \end{example}

% \begin{definition} Let $f=x_{i_2}x_{i_1}$ and $g=x_{j_2}x_{j_1}$ with $\gcd(f,g)=1$ and $i_2>j_2.$ The \emph{degree of collision} of $f$ and $g$ is\\
%     \[\alpha_{{f},{g}}=\left\{
%   \begin{array}{cc}
%        0& i_2>i_1>j_2>j_1  \\
%        1& i_2>j_2>i_1>j_1\\
%        2& i_2>j_2>j_1>i_1
%   \end{array}
%   \right..\]
% \end{definition}

%The condition $\gcd(f,g)$ is utterly important since without it, there is no particular reason for the cardinality of the Minkowski sum to be a power of 2. Take for example $f=g=x_0x_1.$ We have $|\Alow\cdot f|=4$, however, $|\Alow\cdot f+\Alow\cdot g|= 7.$

%Next, we generalise this result to a finite sum of degree two monomials.

%Finally, by demonstrating that there are no collisions for multiplication we deduce the following.
\begin{proposition}\label{pr:cardinal-product-orbits-type2}
    Let $\I$ be a decreasing monomial set with $r=\max_{f\in\I}\deg(f)$ and $\mu\geq 2$ a positive integer. Also, let $f_i\in \I_2$ for $i\in[1,\mu]$ with $\gcd(f_i,f_j)=1$ for any pair $(i,j)\in[1,\mu]\times [1,\mu]$ with $i\neq j.$ Then \vspace{-5pt}
    \begin{multline}
        \left|\Alow_h\cdot h\cdot\sum\limits_{i=1}^{\mu}\Alow\cdot f_i\right|=\\
        \left|\Alow_h\cdot h\right|\times \frac{\prod\limits_{i=1}^{\mu}|\Alow\cdot f_i|}{2^{\sum\limits_{i\neq j}\alpha_{{f_i},{f_j}}}}, 
    \end{multline}
    where given $\ind(f_i)=\{i_1,i_2\}$ and $\ind(f_j)=\{j_1,j_2\}$ we have $\alpha_{f_i,f_j}=2$ if $f_i,f_j$ are not comparable, $\alpha_{f_i,f_j}=1$ if $f_j\preceq f_i, j_2>i_1$, or $\alpha_{f_i,f_j}=0$ if $f_j\preceq f_i, i_1>j_2.$  %we have that $\alpha_{f_i,f_j}$ is the inversion number of the indices of $(i_2,i_1,j_2,j_1).$  $f_i=x_{i_1}x_{i_2},f_j=x_{j_1}x_{j_2}$ %we have $\alpha_{{f},{g}}$ equals $0$ when $i_2>i_1>j_2>j_1$, or $1$ when $i_2>j_2>i_1>j_1$ or $2$ when $i_2>j_2>j_1>i_1.$ 
  %   \[\alpha_{{f},{g}}=\left\{
  % \begin{array}{cc}
  %      0& i_2>i_1>j_2>j_1  \\
  %      1& i_2>j_2>i_1>j_1\\
  %      2& i_2>j_2>j_1>i_1
  % \end{array}
  % \right..\]
\end{proposition}

%Knowing the cardinality of an orbit, we can move to the next step, 
With the cardinality of an orbit at our disposal, we can proceed to enumerate Type II codewords of a given weight. The classification perspective provided by Theorem \ref{thm:equality-orbits-typeII}, which is also constructive, gives the monomials (a pair, triplet, or more monomials) generating such codewords. Additionally, the cardinality of an orbit is determined by Proposition \ref{pr:cardinal-product-orbits-type2}. Lastly, we establish that any two distinct tuples of valid monomials for Type II codewords define disjoint orbits. With these components in place, our formula can be formally stated.   

% \begin{corollary}\label{cor:typeII}
%     Let $\I$ be a decreasing monomial set with $r=\max_{f\in\I}\deg(f).$ Let $\w_{\mu}=2^{m+1-r}-2^{m+1-r-\mu}$ with $2\leq 2\mu \leq m-r+2.$ Then the set of all $\w_{\mu}$-weight codewords of type II is 
%     \begin{multline}
%         W_{\w_{\mu}}=\\ 
%         \bigcup\limits_{\substack{i\in[1,\mu], f_i\in\I_r\\ \forall\; i,j \in [1,\mu], i\neq j, \; h=\gcd(f_i,f_j)\\h \in \I_{r-2}}}\Alow_h\cdot h\cdot\sum_{i=1}^{\mu}\Alow_{f_i}\cdot \frac{f_i}{h}
%     \end{multline}      
% \end{corollary}

\begin{theorem}\label{thm:formula_typeII}Let $\I$ be a decreasing monomial set and $r=\max_{f\in \I}\deg(f).$ Let $\w_{\mu}=2^{m+1-r}-2^{m+1-r-\mu}$ with $2\leq 2\mu \leq m-r+2.$ The number of weight $\w_{\mu}$ codewords of Type II is %$|W_{\w_{\mu}}(\I)|=\sum\limits_{\substack{1\leq i\leq \mu,\;f_i\in \I_r\\ h=\gcd(f_i,f_j) \in \I_{r-2}}} \frac{2^{r-2+2\mu+ |\lambda_h|+\sum\limits_{i=1}^{\mu}|\lambda_{f_i}(\frac{f_i}{h})|}}{2^{\sum\limits_{(f_i,f_j)}\alpha_{\frac{f_i}{h},\frac{f_j}{h}}}}.$
  \begin{equation}\label{eq:formula_typeII}
      |W_{\w_{\mu}}(\I)|= \sum\limits_{\substack{1\leq i\leq \mu,\;f_i\in \I_r\\ h=\gcd(f_i,f_j) \in \I_{r-2}}} \frac{2^{r-2+2\mu+ |\lambda_h|+\sum\limits_{i=1}^{\mu}|\lambda_{f_i}(\frac{f_i}{h})|}}{2^{\sum\limits_{(f_i,f_j)}\alpha_{\frac{f_i}{h},\frac{f_j}{h}}}}.
  \end{equation} 
\end{theorem}

\subsection{Complete weight distribution for subcodes of the second order Reed-Muller code and their duals}

Let us consider decreasing monomial codes $\C(\I)$ within $\R(1,m)\subseteq\C(\I)\subseteq\R(2,m)$. For this case, we have $r=2$ and thus the condition on monomials reduces to $\gcd(f,g)=1$ with $f,g\in \I_2.$ The complete weight distribution for $\R(2,m)$ is already known in terms of orbits under the complete affine group. However, here, the codes will admit a sub-group as permutation group, that is, $\Alow.$ Theorem \ref{thm:formula_typeII} almost provides the complete weight distribution for such codes. Yet, there is one more weight to characterise, namely $2\wm.$ To address this, let us recall a pertinent result. 

\begin{lemma}\cite[Lemma 1]{kasami1970weight}\cite{sloane1970weight}\label{lem:2dmin}
    Let $P\in \Rm$ with $\deg(P)=2.$ Then $\w(\ev(P))=2^{m-1}$ if and only if $P$ is affine equivalent to $x_1x_2+\cdots+x_{2l-1}x_{2l}+x_{2l+1}$ for $0\leq l\leq (m-1)/2$, where $x_i$'s are mutually independent. %\mohammad{I tried to find this result in the provided references but I couldn't. We betetr to give an exact address, LEmma or Theorem number. Also, I wanted to double-check the statement.}   
\end{lemma}

%Using the same techniques, 
This lemma can be similarly expressed in terms of $\Alow.$ Observe that $2\wm$ are thus expressed as linear combinations of weight $2^{m-1}$ codewords and $\wm$-weight codewords. In fact, any polynomial $Q\in\Alow\cdot x_i$ satisfying $\w(\ev(Q))=2^{m-1}$ represents the $x_{2l+1}$ term in Lemma \ref{lem:2dmin}. The remaining task is the evaluation of $P=x_1x_2+\cdots+x_{2l-1}x_{2l}$, which are codewords of weight $2^{m+1-r}-2^{m+1-l}.$ While we know how to determine the number of polynomials of the form $P$, we need to establish how many polynomials of the form $P+Q$ exist within the orbit $\Alow\cdot x_{j}+\sum_{i=1}^{l}\Alow\cdot{f_i}.$ It turns out that the linear form originating from $\Alow\cdot x_{j}$ contributes to the existing orbit with a multiplicative factor equal to $|\Alow_{f_1\dots f_l}\cdot x_j|.$ Consequently, we have the following:%\mohammad{Is this followed by the Theorem?}

\begin{theorem}\label{thm:count-all-subcodes-RM2}
    Let $\C(\I)$ be a decreasing monomial code satisfying $\R(1,m)\subseteq\C(\I)\subseteq \R(2,m).$ The complete weight distribution of $\C(\I)$ is given in Table \ref{tab:weight-distrib-subRM2}.
    \begin{table}[!h]\vspace{-5pt}
        \centering
        \scriptsize
    \begin{tabular}{c|l}
    \toprule
    $|W_{\w}|$& $\w$\\
    \midrule
      1& $\w=0,\w=2^m$\\
      $\sum\limits_{f\in \I_2}2^{2+|\lambda_f|}$& $\w\!=\!2^{m\!-\!2},2^{m}\!-\!2^{m\!-\!2}$\\
      $\sum\limits_{\substack{1\leq i\leq \mu,\;f_i\in \I_2\\ \gcd(f_i,f_j)=1}} \frac{2^{2\mu+ \sum\limits_{i=1}^{\mu}|\lambda_{f_i}|}}{2^{\sum\limits_{(f_i,f_j)}\alpha_{{f_i},{f_j}}}}$& $\w=2^{m\!-\!1}\pm 2^{m\!-\!1\!-\!\mu}$\\
      $\sum\limits_{l=0}^{\frac{m-1}{2}}\;\;\;\;\;\; \smashoperator{\sum\limits_{\substack{1\leq i\leq l,\;f_i\in \I_2\\ \gcd(x_j,f_i)=1\\
      \gcd(f_i,f_j)=1}}} \frac{2^{2\mu+1 \sum\limits_{i=1}^{l}|\lambda_{f_i}|+1+|\lambda_{f_1\dots f_{l}}(x_j)|}}{2^{\sum\limits_{(f_i,f_j)}\alpha_{{f_i},{f_j}}}}$& $\w=2^{m-1}=2\wm$\\
      \bottomrule
      \end{tabular}
        \caption{Complete weight distribution for decreasing monomial code $\C(\I)$ satisfying $\R(1,m)\subseteq\C(\I)\subseteq\R(2,m).$}
        \label{tab:weight-distrib-subRM2}\vspace{-5pt}
    \end{table}
\end{theorem}

Note that the weight distribution of any decreasing monomial code $\C(\I)$ satisfying $\R(1,m)\subseteq\C(\I)\subseteq \R(2,m)$ is symmetric. Indeed, we have $|W_{2^{m}-\w}(\C(\I))|=|W_{\w}(\C(\I))|$ for all possible values of $\w.$

\begin{remark}[High-rate complete weight distribution]
    Using the well-known MacWilliams identities on the weight enumerator polynomial of a code $\C(\I)$ and its dual $\C(\I)^{\bot}$ \cite{Macwil1963}, we can compute the weight distribution of $\C(\I)^{\bot}.$
Let us recall the first equation that relates $|W_i(\C)|$ to $|W_j(\C^{\bot})|.$ 
\begin{equation}
    \sum\limits_{j=0}^{n-\nu}\binom{n-j}{\nu}|W_j(\C)|=2^{k-\nu}\sum\limits_{j=0}^{\nu}\binom{n-j}{n-\nu}|W_j(\C^{\bot})|,
\end{equation}
for $0\leq \nu\leq n.$
    \noindent Furthermore, the following holds: 
    \begin{align*}
        \R(1,m)&\subseteq \C(\I)\subseteq \R(2,m)\\
        \R(m-2,m)&\supseteq \C(\I)^{\bot}\supseteq \R(m-3,m).
    \end{align*}
    Note that the weight distribution of $\C(\I)^{\bot}$ is also symmetric. %\mohammad{Can we add more on the use McWilliams' of identity here?}
\end{remark}
 \begin{example} Let $m=5$ and $\C(\I)$ a code of length $32$, dimension $12$ and minimum distance $8$ defined by %$\I=\{f\in\Mon\mid f\preceq x_0x_4, f\preceq x_1x_3\}$.  We have 
 $\I_2=\{x_0x_1,x_0x_2,x_0x_3,x_0x_4,x_1x_2,x_1x_3\}.$ We have 
 \[\R(1,5)\subseteq \C(\I)\subseteq \R(2,5)\]
 which implies 
 \[\R(2,5)\subseteq \C(\I)^{\bot}\subseteq \R(3,5).\]
 The dual code $\C(\I)^{\bot}$ will be defined by the set $\Mon\setminus \check{\I}.$ Since all degree 1 monomials are in $\I$ and $\check{x_i}=(x_0\cdots x_{4})/x_i$ we have $\check{\I}=\{f\in\Mon\mid \deg(f)=4\}\cup\check{\I_2}.$ Computing $\check{\I_2}=\{x_2x_3x_4,x_1x_3x_4,x_1x_2x_4,x_1x_2x_3,x_0x_3x_4,x_0x_2x_4\}$ gives us the monomials in the dual code, i.e., $\Mon\setminus\check{\I}=\{f\in\Mon\mid \deg(f)\leq 2\}\cup\{x_0x_1x_2,x_0x_1x_3,x_0x_1x_4,x_0x_2x_3\}.$ 
 We have %The weight distribution of $\C(\I)$ and its dual is
 \begin{equation*}
     W(\I,X)=1+108X^8+576X^{12}+2726X^{16}.\vspace{-10pt}
 \end{equation*}

These results are obtained by our formulae as follows.
 \begin{itemize}
     \item $\w=8:$ For $f\in \I_2$ we have $\lambda_f\in\{(0,0),(0,1),(0,2),\\(0,3),(1,1),(1,2)\}$ which makes $|\lambda_f|\in \{0,1,2,3,2,3\}.$ Hence, we have $2^2\times(1+2+4+8+4+8)=4\times 27=108.$
     \item $\w=12:$ The valid pairs $(f,g)$ with $\gcd(f,g)=1$ are $\{(x_0x_2,x_1x_3),(x_0x_3,x_1x_2),(x_0x_4,x_1x_2),(x_0x_4,x_1x_3).$ Computing for each pair the values $[|\lambda_f|,|\lambda_g|,\alpha_{f,g}]$ gives $\{[1,3,1],[2,2,2],[3,2,2],[3,3,2]\}$ which gives $|\lambda_f|+|\lambda_g|-\alpha_{f,g}$ equal to $\{3,2,3,4\}.$ Finally, we obtain a total of $2^{4}\times(2^3+2^2+2^3+2^4)=16\times 36=576.$
     \item $\w=16:$ The valid tuples $(f,g,x_j)$ are $\{(x_0x_2,x_1x_3,x_4),(x_0x_3,x_1x_2,x_4),(x_0x_4,x_1x_2,x_3),\\(x_0x_4,x_1x_3,x_2).$ For each element $x_j$ in the tuples its $|\lambda_{fg}(x_j)|$ is equal to $1.$ In conclusion we have $2^{1}\times 2^{1}\times 572=2726$ codewords of weight $16.$ 
 \end{itemize}
 For the dual code we obtain
\begin{multline*}
     W(\Mon\setminus\check{\I},X)=1+88X^4+ 128X^{6}+ 5596X^{8}+ 30336X^{10}\\+ 116072X^{12}+ 215296X^{14}+ 313542X^{16}. 
\end{multline*}
Let us detail the computation of the first terms in the weight distribution of the dual code, using the MacWilliams identities. 
\begin{itemize}
    \item $\w=0:$ \begin{align*}
    \sum\limits_{j=0}^{n}|W_j(\C)|&=2^{k}|W_0(\C^{\bot})|\\
    2^{12}&=2^{12}|W_0(\C^{\bot})|
\end{align*}
    \item $\w=1:$\begin{align*}
    \sum\limits_{j=0}^{n-1}(n-1)|W_j(\C)|&=2^{k-1}(n|W_0(\C^{\bot})|+|W_1(\C^{\bot})|)\\
    2^{16}&=2^{11}(2^5+|W_0(\C^{\bot})|).
\end{align*} 
    \item $\w=2:$\begin{align*}
    \sum\limits_{j=0}^{n-2}\binom{n-j}{2}|W_j(\C)|&=2^{k-2}(\binom{n}{2}|W_0(\C^{\bot})|+|W_2(\C^{\bot})|)\\
    507904&=2^{10}(496+|W_2(\C^{\bot})|).
\end{align*} 
    \item $\w=3:$\begin{align*}
    \sum\limits_{j=0}^{n-3}\binom{n-j}{3}|W_j(\C)|&=2^{k-3}(\binom{n}{3}|W_0(\C^{\bot})|+|W_3(\C^{\bot})|)\\
    2539520&=2^{9}(4960+|W_3(\C^{\bot})|).
\end{align*}  
    \item $\w=4:$\begin{align*}
    \sum\limits_{j=0}^{n-4}\binom{n-j}{4}|W_j(\C)|&=2^{k-4}(\binom{n}{4}|W_0(\C^{\bot})|+|W_4(\C^{\bot})|)\\
    9228288&=2^{8}(35960+|W_4(\C^{\bot})|)\\
    36048&=35960+|W_4(\C^{\bot})|.
\end{align*} 

\end{itemize}

 \end{example}

\subsection{Numerical results}

We have implemented our formulae and verified them for all the known Reed-Muller codes from \cite{weightRM-link}. Table \ref{tab:wt_distrib_m7_9} lists the results of the enumeration of two example codes; a polar code constructed solely based on the reliability rule; that is, $|\I^{\text{Polar}}|\!=\!K$, and RMxPolar codes \cite[Section VI.1]{rowshan-pac1} constructed using both the reliability rule and the weight rule; that is, $\I\!=\!\I^{\text{RM}}\!\cup\!\I^{\text{Polar}}$. 
%\end{equation*}
To construct the RMxPolar code ($2^m$,$K$), we find $r'=\arg\max_{r} \sum_{j=0}^{r}{m \choose j}\leq K$. Then we collect all monomials of degree up to ${r'}$ in set $\I^{\text{RM}}$; hence $|\I^{\text{RM}}|=K'=\sum_{j=0}^{r'}{m \choose j}$. The rest are chosen among the monomials of degree ${r'+1}$ with the highest reliability and are collected in the set $\I^{\text{Polar}}$ with $|\I^{\text{Polar}}|=K-K'$. Note that when $K=K'$, we have a Reed-Muller code. Given the intersection with $\R(r',m)$, this code is called the crossed RM-Polar code, and is hence denoted RMxPolar. %Whereas for polar codes, all monomials are chosen based on the reliability rule; hence, $|\I^{\text{Polar}}|=K$. 
Observe that both codes are considered decreasing monomial codes, and our formulae can be used for.

% \begin{center}
% \setlength{\tabcolsep}{1pt}
% \footnotesize
% \begin{tabular}{ l|c|c c c c  } 
%   & $\wm$ & $|W_{\wm}|$ & $|W_{1.5\wm}|$ & $|W_{1.75\wm}|$ & $|W_{2\wm}|$ \\ 
%   \hline
%  RMxP(128,25) & 32 & x &  &  &  &  \\ 
%  $P(128,25)$ & 16 & x &  &  &  &  \\ 
%  \hline
%  $RMxP(256,30)$ & 64 & x &  &  &  &  \\ 
%  $P(256,30)$ & 32 & x &  &  &  &  \\ 
%  \hline
%  $RMxP(512,40)$ & 128 & x &  &  &  &  \\ 
%  $P(512,40)$ & 64 & x &  &  &  &  \\ 
%  \hline
% \end{tabular}
% \end{center}

\setlength{\tabcolsep}{3pt}
\begin{table}[!ht]
    \centering
    \begin{tabular}{rl|r|r||rl|r|r}\toprule
     %\multirow{2}{*}{$\w$} 
     \multicolumn{4}{c}{(128,25)} & \multicolumn{4}{c}{(256,30)}\\
     \cmidrule(lr){1-4} \cmidrule(lr){5-8}
    %\toprule
     \multicolumn{2}{c|}{$\w$} &  RMxPolar & Polar & \multicolumn{2}{c|}{$\w$} &  RMxPolar & Polar\\
        \midrule
        0& 128 & 1 & 1 & 0& 256  &  1  &  1 \\
       {\color{blue}16}& 112 & 0 & {\color{blue}24} & {\color{blue}32}& 224  &  0  &  {\color{blue}88} \\
       {\color{blue}32}& 96 & {\color{blue}2476} & 4444 & 48& 208  &  0  &  128 \\
       40& 88 & 0 & 10240 & {\color{blue}64}& 192  &  {\color{blue}5292} & 16860 \\
       48& 80 & 474432 & 633256 & 80& 176  & 0  & 177792  \\
       56& 72 & 6451200 & 5691392 & 96& 160 & 2302272 & 6444392 \\
        \multicolumn{2}{c|}{64} & 19698214 & 20875718 & 112& 144 &  86388736 & 89786624 \\
        & & & & 120& 136 & 154140672 & 29360128 \\
         & & & & \multicolumn{2}{c|}{128} & 588067878 & 813781190\\
         \bottomrule
    \end{tabular}
    \caption{Weight distribution for (128,25) and (256,30) codes. }
    \label{tab:wt_distrib_m7_9}
\end{table}
Note that Table \ref{tab:weight-distrib-subRM2} can be used for the weight distribution of RMxPolar codes provided in Table \ref{tab:wt_distrib_m7_9}, where $\C(\I)\subset\R(2,m)$. However, in the case of the polar codes in Table \ref{tab:wt_distrib_m7_9}, we have $\C(\I)\subset\R(3,m)$. This can be realized from $\wm=2^{m-3}$ of each code; $\wm=16=2^{7-3}$ and $\wm=32=2^{8-3}$ for polar codes of (128,25) and (256,30), respectively. For the weight distribution of polar codes we have the formula for $\wm$ and $1.5\wm$ from \cite{bardet,rowshan1.5w}, the remaining weights being computed using MAGMA's Leon Algorithm \cite{magma}.  

%%%%%%%%%%%%%%%%%%%%%%%%%%%%%%%%%%%%%%%%%%%%%%%%%%%%%%%%%%%%%

\begin{figure}[!ht]
%    \centering
\centering\resizebox{\columnwidth}{!}{
\begin{tikzpicture}[scale=2,thick]%, background rectangle/.style={fill=olive!45}, show background rectangle]

% \node at (1,-2) (0) {$\bm{x_0}$};
% \node at (1,-1) (1) {$\bm{x_1}$};
% \node at (1,0) (2) {$\bm{x_2}$};
% \node at (1,1) (3) {$\bm{x_3}$};
% \node at (1,2) (4) {$\bm{x_4}$};
% \node at (1,3) (5) {$\bm{x_5}$};
% \node at (1,4) (6) {$\bm{x_6}$};
\node at (0,0) (01) {\Large$\color{brown}\bm{x_0x_1}$};

\node at (1,0) (l1) {\Large$\color{violet} \bm{0}$};

\node at (0,1) (02) {\Large$\color{brown}\bm{x_0x_2}$};

\node at (1,1) (l2) {\Large$\color{violet}\bm{ 1}$};

\node at (0,2) (03) {\Large$\color{brown}\bm{x_0x_3}$};

\node at (1,2) (l3) {\Large$\color{violet}\bm{ 2}$};

\node at (0,3) (04) {\Large$\color{brown}\bm{x_0x_4}$};

\node at (1,3) (l4) {\Large$\color{violet}\bm{ 3}$};

\node at (0,4) (05) {\Large$\color{brown}\bm{x_0x_5}$};

\node at (1,4) (l5) {\Large$\color{violet}\bm{ 4}$};

%\node at (0,5) (06) {$\bm{x_0x_6}$};
\node at (-1,2) (12) {\Large$\color{brown}\bm{x_1x_2}$};
\node at (-1,3) (13) {\Large$\color{brown}\bm{x_1x_3}$};
\node at (-1,4) (14) {\Large$\color{brown}\bm{x_1x_4}$};
\node at (-1,5) (15) {$\bm{x_1x_5}$};

\node at (0,5) (l15) {\Large$\color{violet}\bm{ 5}$};

%\node at (-1,6) (16) {$\bm{x_1x_6}$};
\node at (-2,4) (23) {\Large$\color{brown}\bm{x_2x_3}$};
\node at (-2,5) (24) {$\bm{x_2x_4}$};
\node at (-2,6) (25) {$\bm{x_2x_5}$};

\node at (-1,6) (l25) {\Large$\color{violet}\bm{ 6}$};

%\node at (-2,7) (26) {$\bm{x_2x_6}$};
\node at (-3,6) (34) {$\bm{x_3x_4}$};
\node at (-3,7) (35) {$\bm{x_3x_5}$};

\node at (-2,7) (l35) {\Large$\color{violet}\bm{ 7}$};

%\node at (-3,8) (36) {$\bm{x_3x_6}$};
\node at (-4,8) (45) {$\bm{x_4x_5}$};

\node at (-3,8) (l45) {\Large$\color{violet}\bm{ 8}$};

%\node at (-4,9) (46) {$\bm{x_4x_6}$};
%\node at (-5,10) (56) {$\bm{x_5x_6}$};

\draw[line width=0.25mm](01) --(02) --(03) -- (04) --(05);
\draw[line width=0.25mm](02) --(12) --(13) -- (14) -- (15);
\draw[line width=0.25mm](03) --(13) --(23) -- (24) -- (25);
\draw[line width=0.25mm](04) --(14) --(24) -- (34) -- (35);
\draw[line width=0.25mm](05) --(15) --(25) -- (35) -- (45);

\draw[line width=0.25mm,violet](12) --(04) --(23) -- (15) -- (34);
\draw[line width=0.25mm,violet](13) --(05) --(24) ;

\node at (0-5,0) (01b) {\Large$\color{brown}\bm{x_0x_1}$};
\node at (0-5,1) (02b) {\Large$\color{brown}\bm{x_0x_2}$};
\node at (0-5,2) (03b) {\Large$\color{brown}\bm{x_0x_3}$};
\node at (0-5,3) (04b) {\Large$\color{brown}\bm{x_0x_4}$};
\node at (0-5,4) (05b) {\Large$\color{brown}\bm{x_0x_5}$};
%\node at (0,5) (06) {$\bm{x_0x_6}$};
\node at (-1-5,2) (12b) {\Large$\color{brown}\bm{x_1x_2}$};
\node at (-1-5,3) (13b) {\Large$\color{brown}\bm{x_1x_3}$};
\node at (-1-5,4) (14b) {\Large$\color{brown}\bm{x_1x_4}$};
\node at (-1-5,5) (15b) {$\bm{x_1x_5}$};
%\node at (-1,6) (16) {$\bm{x_1x_6}$};
\node at (-2-5,4) (23b) {\Large$\color{brown}\bm{x_2x_3}$};
\node at (-2-5,5) (24b) {$\bm{x_2x_4}$};
\node at (-2-5,6) (25b) {$\bm{x_2x_5}$};
%\node at (-2,7) (26) {$\bm{x_2x_6}$};
\node at (-3-5,6) (34b) {$\bm{x_3x_4}$};
\node at (-3-5,7) (35b) {$\bm{x_3x_5}$};
%\node at (-3,8) (36) {$\bm{x_3x_6}$};
\node at (-4-5,8) (45b) {$\bm{x_4x_5}$};
%\node at (-4,9) (46) {$\bm{x_4x_6}$};
%\node at (-5,10) (56) {$\bm{x_5x_6}$};

\draw[line width=0.25mm](01b) --(02b) --(03b) -- (04b) --(05b);
\draw[line width=0.25mm](02b) --(12b) --(13b) -- (14b) -- (15b);
\draw[line width=0.25mm](03b) --(13b) --(23b) -- (24b) -- (25b);
\draw[line width=0.25mm](04b) --(14b) --(24b) -- (34b) -- (35b);
\draw[line width=0.25mm](05b) --(15b) --(25b) -- (35b) -- (45b);

\draw[line width=0.25mm,violet](12b) --(04b) --(23b) -- (15b) -- (34b);
\draw[line width=0.25mm,violet](13b) --(05b) --(24b) ;

% \node at (0-5-5,0) (01c) {$\bm{x_0x_1}$};
% \node at (0-5-5,1) (02c) {$\bm{x_0x_2}$};
% \node at (0-5-5,2) (03c) {$\bm{x_0x_3}$};
% \node at (0-5-5,3) (04c) {$\bm{x_0x_4}$};
% \node at (0-5-5,4) (05c) {$\bm{x_0x_5}$};
% %\node at (0,5) (06) {$\bm{x_0x_6}$};
% \node at (-1-5-5,2) (12c) {$\substack{\bm{x_1x_2}\\\bm{3}}$};
% \node at (-1-5-5,3) (13c) {$\bm{x_1x_3}$};
% \node at (-1-5-5,4) (14c) {$\bm{x_1x_4}$};
% \node at (-1-5-5,5) (15c) {$\bm{x_1x_5}$};
% %\node at (-1,6) (16) {$\bm{x_1x_6}$};
% \node at (-2-5-5,4) (23c) {$\bm{x_2x_3}$};
% \node at (-2-5-5,5) (24c) {$\bm{x_2x_4}$};
% \node at (-2-5-5,6) (25c) {$\bm{x_2x_5}$};
% %\node at (-2,7) (26) {$\bm{x_2x_6}$};
% \node at (-3-5-5,6) (34c) {$\bm{x_3x_4}$};
% \node at (-3-5-5,7) (35c) {$\bm{x_3x_5}$};
% %\node at (-3,8) (36) {$\bm{x_3x_6}$};
% \node at (-4-5-5,8) (45c) {$\bm{x_4x_5}$};
% %\node at (-4,9) (46) {$\bm{x_4x_6}$};
% %\node at (-5,10) (56) {$\bm{x_5x_6}$};

% \draw[line width=0.25mm](01c) --(02c) --(03c) -- (04c) --(05c);
% \draw[line width=0.25mm](02c) --(12c) --(13c) -- (14c) -- (15c);
% \draw[line width=0.25mm](03c) --(13c) --(23c) -- (24c) -- (25c);
% \draw[line width=0.25mm](04c) --(14c) --(24c) -- (34c) -- (35c);
% \draw[line width=0.25mm](05c) --(15c) --(25c) -- (35c) -- (45c);

\foreach \i in {13,14} 
 \draw [{Stealth[length=3mm]}-{Stealth[length=3mm]},thick,red] (02b) to [bend right]  (\i);
 
% \foreach \i in {01,14} 
%  \draw [{Stealth[length=3mm]}-{Stealth[length=3mm]},thick,red] (23b) to [bend right]  (\i);
\draw [{Stealth[length=3mm]}-{Stealth[length=3mm]},thick,red] (01b) to [bend right]  (23);
\draw [{Stealth[length=3mm]}-{Stealth[length=3mm]},thick,red] (14b) to [bend right]  (23);

\foreach \i in {14} 
 \draw [{Stealth[length=3mm]}-{Stealth[length=3mm]},thick,red] (03b) to [bend right]  (\i);

\foreach \i in {12,13,23} 
 \draw [{Stealth[length=3mm]}-{Stealth[length=3mm]},thick,red] (04b) to [bend right]  (\i);

\foreach \i in {12,14,13,23} 
 \draw [{Stealth[length=3mm]}-{Stealth[length=3mm]},thick,red] (05b) to [bend right]  (\i);

%\draw [{Stealth[length=3mm]}-{Stealth[length=3mm]},thick,blue] (15) to [bend right]  (04b) to  [bend right]  (23);

% \draw [{Stealth[length=3mm]}-{Stealth[length=3mm]},thick,blue] (15b) to [bend left]  (23) to  [bend left]  (04b);

\draw [{Stealth[length=3mm]}-{Stealth[length=3mm]},thick,blue] (23) to [bend right]  (14b) to  [bend right]  (05);
% \draw [{Stealth[length=3mm]}-{Stealth[length=3mm]},thick,blue] (23b) to [bend left]  (14) to  [bend left]  (05b);

% \foreach \i in {01,04,05,14,15,45} 
%  \draw [{Stealth[length=3mm]}-{Stealth[length=3mm]},thick] (23c) to [bend left]  (\i b);

\draw[line width=0.25mm, loosely dotted,  violet] (01) -- (l1);

\draw[line width=0.25mm, loosely dotted,  violet] (02) -- (l2);

\draw[line width=0.25mm, loosely dotted,  violet] (12) -- (03) -- (l3);

\draw[line width=0.25mm, loosely dotted,  violet] (13) -- (04) -- (l4);

\draw[line width=0.25mm, loosely dotted,  violet]  (23) -- (14) -- (05) -- (l5);

\draw[line width=0.25mm, loosely dotted,  violet] (24) -- (15) -- (l15);

\draw[line width=0.25mm, loosely dotted,  violet] (34) -- (25) -- (l25);

\draw[line width=0.25mm, loosely dotted,  violet] (35) -- (l35);

\draw[line width=0.25mm, loosely dotted,  violet] (45) -- (l45);

\end{tikzpicture}
}
    \caption{Non-zero codewords creation for $m=6.$ Monomials in $\I$ (bold brown) are defined by $f\preceq x_1x_4,f\preceq x_0x_5$ and $f\preceq x_2x_4.$ For $\wm$ each level gives $|\lambda_f|$, for $1.5\wm$ each red arrow indicate $(f,g)$ with $\gcd(f,g)=1$, whereas for $1.75\wm$ blue arrows indicate triplets $(f,g,h)$ with $\gcd(f,g)=\gcd(g,h)=1.$}
    \label{fig:codwords_formation_m7}
\end{figure}
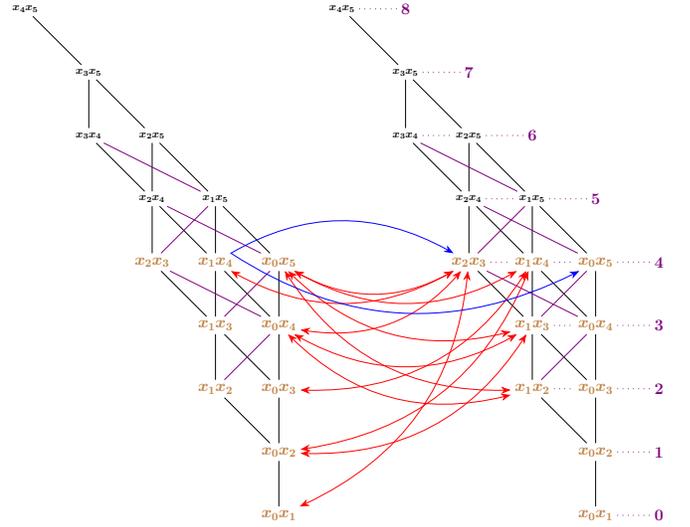

\section{A partial order based on weight contribution}

%\vlad{Give the $m=7$ example to motivate.}%\mohammad{Is it still partial order? I believe the order is exact. Also, this order may overlap with paths on partial order graph (but the overal shape of graph will be different, I will draw an example I have in mind and send it to you later today.) because as if we consider the lef -wsap in Scurch paper as a way to increase the relaibility, that is a way to reduce the contribution or the cardinality of the orbit (which is what we want), hence the more relaible the coordinate/monomial/channel, the lower the contribution. Nevertheless, it is worth introducing it.}

%\vlad{Indeed, the more reliable the channel is, the less is its contribution to the minimum weight codewords. However, the construction problem is not to decide between comparable channels/monomials, but to decide between non-comparable (see Mondelli's paper on construction and the EXample I gave in the Appendix for $m=6.$) Nevertheless, I am working on the properties of the restriction of the poset on $\I_2$ and this is a much simpler case that the complete poset. Therefore, its combination with weight contribution should not be that hard! For the moment we have a graph representation of the possible connections for different values of $\mu$, which helps a lot. It would be great if we could find more mathematical/structural properties to reduce the complexity.}
It is widely recognized that there are partial orders for channel reliability \cite{mori,bardet,schurch,Wang-Dragoi2023,Kahraman2017,Ordentlich-Roth2017,Wu-Siegel2019,Dragoi-CristescuPO2021}. However, to the best of our knowledge, there is no partial order on the weight contribution. In this section, we introduce a new partial order relation with respect to weight contribution. We use the symbol $\preccurlyeq$ instead of $\preceq$ 
 in Definition \ref{def:order} to distinguish it from the partial order of reliability. First, we define this partial order relation for decreasing monomial codes with respect to weight distribution, and then we extend it to the order relation of monomials in the generating set of decreasing monomial codes with respect to their weight contribution. 
\begin{definition}
    Let $\C(\I), \C(\J)$ be two decreasing monomial codes. Let $\w\in[0,2^{m}]$ and define %The order relation $\preccurlyeq_{\w}$ will be defined with respect to a weight $\w\in[0,2^{m}].$
    \[\C(\I)\preccurlyeq_{\w} \C(\J)\text{ if } |W_{\w}(\J)|\leq |W_{\w}(\I)|.\] 

    We say that $\C(\I)\preccurlyeq_{[\w_1,\w_2]}\C(\J)$ if $\forall \w\in [\w_1,\w_2]$ we have $\C(\I)\preccurlyeq_{\w} \C(\J).$ 
\end{definition}
Unless $|\I|\neq |\J|$, we cannot have $\C(\I)\preccurlyeq_{[1,2^m]}\C(\J).$ In the case of $|\I|=|\J|$, it is thus natural to examine the smallest interval for $w$ on which two codes are comparable. Having determined the structure of codewords of Type II for $\w\in [\wm,2\wm)$, we can now establish the order relation $\preccurlyeq_{[\wm,2\wm)}.$ While our construction and definitions are applicable to any dimension, for consistency with our results on subcodes of $\R(2,m)$ where we have the complete weight distribution, we shall consider this particular case. 
We extend the definition $\preccurlyeq_{\wm}$ to monomials as follows.
\begin{definition}
    Let $r$ be the maximum degree of monomials and $f,g\in\I_r$. Then, we have $f\preccurlyeq_{\wm} g$ if and only if $|\lambda_f|<|\lambda_g|$.
\end{definition}

%In particular we will look at the low-rate decreasing monomial codes. 

%\vlad{Take $m=7.$ CRC specification: start with a polar of dimension 12 or 19 which becomes a code of dimension 21 or 28. The CRC-RM-Polar will do the same on a RM-Polar code of dimension 12 or 19. Check these two particular dimensions.}

To each degree $r$, one can associate the partial order set (poset) of $\I_r$ defined by $\preceq$ and $\preccurlyeq_{\wm}.$ 
In Fig. \ref{fig:codwords_formation_m7} we illustrate the poset for $m=7$ (black lines for $\preceq$, violet lines for $\preccurlyeq_{\wm}$).

\begin{figure}[!ht]
%    \centering
\centering\resizebox{\columnwidth}{!}{
\begin{tikzpicture}[scale=2,thick]%, background rectangle/.style={fill=olive!45}, show background rectangle]

\node at (0,0) (01) {$0$};
\node at (-1,0) (01s) {\Large$\color{violet}S_0=1$};

\node at (0,1) (02) {$1$};
\node at (-1.5,1) (02s) {\Large$\color{violet}S_1=1$};

\node at (0,2) (03) {$2$};
\node at (-2,2) (03s) {\Large$\color{violet}S_2=2$};

\node at (0,3) (04) {$3$};
\node at (-2.5,3) (04s) {\Large$\color{violet}S_3=2$};

\node at (0,4) (05) {$4$};
\node at (-3,4) (05s) {\Large$\color{violet}S_4=3$};

\node at (0,5) (06) {$5$};
\node at (-3.5,5) (06s) {\Large$\color{violet}S_5=3$};

\node at (-1,2) (12) {$2$};
\node at (-1,3) (13) {$3$};
\node at (-1,4) (14) {$4$};
\node at (-1,5) (15) {$5$};
\node at (-1,6) (16) {$6$};
\node at (-4,6) (16s) {\Large$\color{violet}S_6=3$};

\node at (-2,4) (23) {$4$};
\node at (-2,5) (24) {$5$};
\node at (-2,6) (25) {$6$};
\node at (-2,7) (26) {$7$};
\node at (-4.5,7) (26s) {\Large$\color{violet}S_7=2$};

\node at (-3,6) (34) {$6$};
\node at (-3,7) (35) {$7$};
\node at (-3,8) (36) {$8$};
\node at (-5,8) (36s) {\Large$\color{violet}S_8=2$};

\node at (-4,8) (45) {$8$};
\node at (-4,9) (46) {$9$};
\node at (-5.5,9) (46s) {\Large$\color{violet}S_9=1$};

\node at (-5,10) (56) {$10$};
\node at (-6,10) (56s) {\Large$\color{violet}S_{10}=1$};

\draw[line width=0.25mm](01) --(02) --(03) -- (04) --(05) --(06) -- (16) -- (26) -- (36) -- (46) -- (56);
\draw[line width=0.25mm](02) --(12) --(13) -- (14) -- (15) -- (16);
\draw[line width=0.25mm](03) --(13) --(23) -- (24) -- (25) -- (26);
\draw[line width=0.25mm](04) --(14) --(24) -- (34) -- (35) -- (36);
\draw[line width=0.25mm](05) --(15) --(25) -- (35) -- (45) -- (46);

\draw[line width=0.25mm, loosely dotted,  violet] (01) -- (01s);
\draw[line width=0.25mm, loosely dotted,  violet] (02) -- (02s);
\draw[line width=0.25mm, loosely dotted,  violet] (03) -- (03s);
\draw[line width=0.25mm, loosely dotted,  violet] (04) -- (04s);
\draw[line width=0.25mm, loosely dotted,  violet] (05) -- (05s);
\draw[line width=0.25mm, loosely dotted,  violet] (06) -- (06s);
\draw[line width=0.25mm, loosely dotted,  violet] (16) -- (16s);
\draw[line width=0.25mm, loosely dotted,  violet] (26) -- (26s);
\draw[line width=0.25mm, loosely dotted,  violet] (36) -- (36s);
\draw[line width=0.25mm, loosely dotted,  violet] (46) -- (46s);
\draw[line width=0.25mm, loosely dotted,  violet] (56) -- (56s);

\end{tikzpicture}
}
    \caption{A diagram illustrating the sequence of  $|\lambda_f|$ for all $f\in \I_2$ and $m=7.$ The maximum value belongs to $x_5x_6$, which gives $|\lambda_{x_5x_6}|=2(m-2)=10.$ The sequence of $S_j$ is symmetric: $1,1,2,2,3,3,3,2,2,1,1.$}
    \label{fig:lambdaf_seq}
\end{figure}
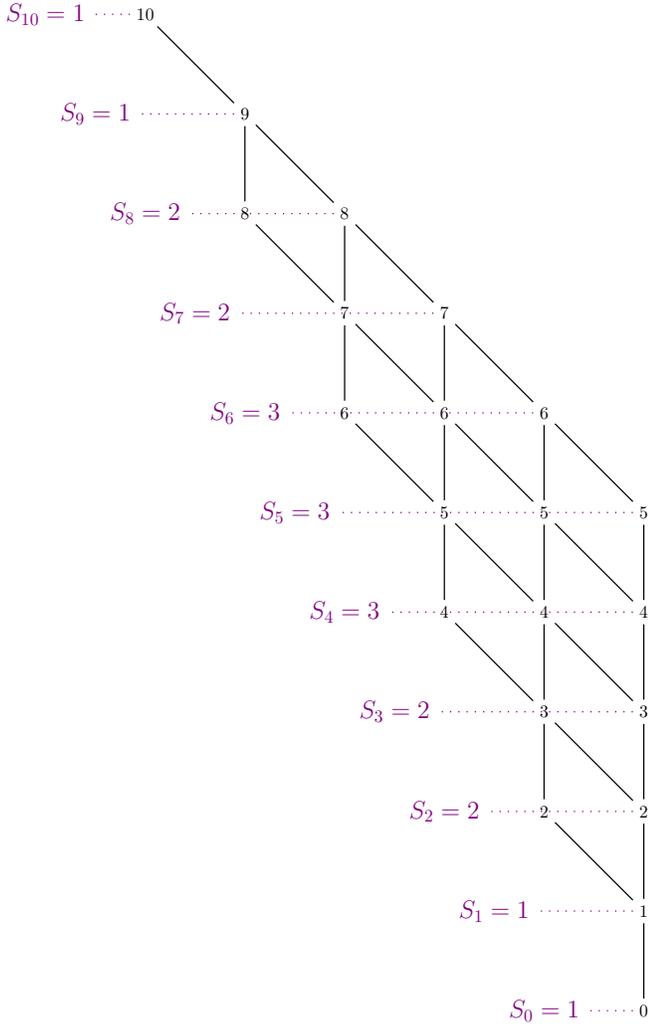

%\subsection{Properties}
    \subsection{Symmetric property and non-comparable elements}
    The order $\preccurlyeq_{\wm}$ adds more restrictions and hence reduces the number of non-comparable elements. To be more precise, when considering monomials of degree $2$, there are $\lfloor\frac{j+2}{2}\rfloor$ monomials at level $j$ and $m-2+j$ with $j\in[0,m-2].$ Defining $S_j=\{f\in\I_2\mid |\lambda_f|=j\}$, we have $|S_j|=\lfloor\frac{j+2}{2}\rfloor$, which can be demonstrated in the following lemma. 
\begin{lemma}\label{lem:poset_properties}
    Let $S_l=\{f\in\I_2\mid |\lambda_f|=l\}.$ Then we have 
    \begin{itemize}
        \item Symmetry: $|S_l|=|S_{m-2+l}|$ for all  $l\in [0,m-2].$
        \item Cardinality: $|S_l|=\lfloor\frac{l+2}{2}\rfloor$ for any $l\in [0,m-2].$
    \end{itemize}
\end{lemma}

\begin{IEEEproof}
    Using Definition \ref{def:lambda_f} for $\lambda_f$, each level $l$ is given by monomials with indices in the subsets of the set $\{\{0,1\},\dots \{m-2,m-1\}\}$ having the same sum. More exactly, we have $S_l=\{x_0x_{l+1},x_1x_{l},\dots x_{\lfloor\frac{l}{2}\rfloor}x_{\lfloor\frac{l+2}{2}\rfloor}\}.$ Indeed, $|\lambda_{x_0x_{l+1}}|=\dots = |\lambda_{x_{\lfloor\frac{l}{2}\rfloor}x_{\lfloor\frac{l+2}{2}\rfloor}}|=l.$ There are ${\lfloor\frac{l+2}{2}\rfloor}$ elements in the set $S_l$, which finishes our proof. 
\end{IEEEproof}
   
    Hence, according to Lemma \ref{lem:poset_properties}, the sequence $S_j$ is not only symmetric, but also uni-modal, with a maximum at $j=m-2$ where we have $|S_{m-2}|=\lfloor\frac{m}{2}\rfloor.$ For example, for $m=6$ the sequence $1,1,2,2,3,2,2,1,1$ gives $|\{f\in\I_2\mid |\lambda_f|=j\}|.$ In Fig. \ref{fig:lambdaf_seq}, we illustrate the symmetry and uni-modality for the case $m=7.$ 
   
%   \textbf{Duality:} Let $\C(\I),\C(J)$ be two decreasing monomial codes with equal dimension satisfying $\R(1,m)\subseteq \C \subseteq \R(2,m).$ Then 
 %   \[\C(\I)\preccurlyeq_{\wm}\C(\J)\Leftrightarrow \C(I)^{\bot}\preccurlyeq_{\wm}\C(\J)^{\bot}.\]

\subsection{Ordering procedure}

There are several possible combinations of the aforementioned orders, resulting in possibly different codes. Due to space constraints, we will briefly propose here one combination, directly applied on the subcodes of $\R(2,m).$ So, let us suppose that we need to construct a code of dimension $1+m< K < 1+m+\binom{m}{2}$ that contains $\R(1,m).$ Our algorithm proceeds in the following way.
\begin{enumerate}
    \item Based on $\preccurlyeq_{\wm}$, select the monomials on the first $l-1$ levels, where
    $1+m+\sum\limits_{j=0}^{l-1}\lfloor\frac{j+2}{2}\rfloor\leq K < 1+m+\sum\limits_{j=0}^{l}\lfloor\frac{j+2}{2}\rfloor.$
    \item The remaining monomials, all from level $l$ will be selected based on the reliability rule, where the reliability can be calculated using different methods such has beta-expansion \cite{he}.  
\end{enumerate}

\begin{example}\label{ex:5}
    Let us take $m=6$ and suppose we need a code of dimension $15.$ Since we have $1+6+(1+1+2+2)<15<1+6+(1+1+2+2+3)$, it implies that $l=4$; hence two more monomials are to be selected. On the level $l=4$ in Fig. \ref{fig:codwords_formation_m7}, we see that there are 3 non-comparable elements, given by our formula $\lfloor\frac{4+2}{2}\rfloor=3$, namely $S_4=\{x_2x_3,x_1x_4,x_0x_5\}.$   
    Equivalently, if we let $\I_2=\{f\mid f\preceq g, g \in S_4\}$, we need to subtract one monomial from $S_4$ in order to obtain the desired code. %In Fig. \ref{fig:codwords_formation_m7}, we illustrate how to compute the weight distribution based on a graph representation. 
 The weight polynomial using our formula for $\C(\I)$ is 
\[W(\I,X)=1+ 300X^{16}+ 5952X^{24}+ 4096X^{28}+ 44838X^{32}.\]

%\[W(\I,X)=1+428X^{16}+11584X^{24}+12288X^{28}+ 82470X^{32}.\]%+ 12288x^{36}+
%11584x^{40}+ 428x^{48} x^{64}.\]
%Let us first look at the minimum weight codewords, the second term in $W(\I,X)$ above. Following \eqref{eq:A_wm} and \eqref{eq:sum_A_wm}, as illustrated in Fig. \ref{fig:codwords_formation_m7}, we have 3 elements at level $|\lambda_f|=4$ (which gives $2^{|\lambda_f|}=2^4$), plus 2 elements at level 3 ($2\times 2^3$), plus 2 at level 2 ($2\times 2^2$), plus one element at level 1 ($2^1$), plus one element at level 0 ($2^0$), which gives a total of $\sum_{f\in\I_r} 2^{|\lambda_f|} = 3\times 16+2\times 8+2\times 2+2+ 1= 75.$ Multiplying this by $2^{\deg(f)}=2^2$ (representing the translations, denoted by $\bve$ in Section \ref{ssec:perm_grp}) gives $\sum_f 2^{\deg(f)+|\lambda_f|}=300.$ 
Removing any element $f\in S_4$ from $\I$ gives the same weight distribution
 \[W(\I\setminus\{f\}),X)=1+ 236X^{16}+ 3136X^{24}+ 26022X^{32}\]

% \[W(\I\setminus\{x_1x_4\}),X)=1+ 236X^{16}+ 3136X^{24}+ 26022X^{32}\]
 
%  \[W(\I\setminus\{x_2x_3\}),X)=1+ 236X^{16}+ 3136X^{24}+ 26022X^{32}\]

\end{example}

\begin{figure}[!ht]
%    \centering
\centering\resizebox{\columnwidth}{!}{
\begin{tikzpicture}[scale=2,thick]%, background rectangle/.style={fill=olive!45}, show background rectangle]

\node at (1,-2) (0) {$\bm{x_0}$};
\node at (1,-1) (1) {$\bm{x_1}$};
\node at (1,0) (2) {$\bm{x_2}$};
\node at (1,1) (3) {$\bm{x_3}$};
\node at (1,2) (4) {$\bm{x_4}$};
\node at (1,3) (5) {$\bm{x_5}$};
\node at (1,4) (6) {$\bm{x_6}$};
\node at (0,0) (01) {$\bm{x_0x_1}$};
\node at (0,1) (02) {$\bm{x_0x_2}$};
\node at (0,2) (03) {$\bm{x_0x_3}$};
\node at (0,3) (04) {$\bm{x_0x_4}$};
\node at (0,4) (05) {$\bm{x_0x_5}$};
\node at (0,5) (06) {$\bm{x_0x_6}$};
\node at (-1,2) (12) {$\bm{x_1x_2}$};
\node at (-1,3) (13) {$\bm{x_1x_3}$};
\node at (-1,4) (14) {$\bm{x_1x_4}$};
\node at (-1,5) (15) {$\bm{x_1x_5}$};
\node at (-1,6) (16) {$\bm{x_1x_6}$};
\node at (-2,4) (23) {$\bm{x_2x_3}$};
\node at (-2,5) (24) {$\bm{x_2x_4}$};
\node at (-2,6) (25) {$\bm{x_2x_5}$};
\node at (-2,7) (26) {$\bm{x_2x_6}$};
\node at (-3,6) (34) {$\bm{x_3x_4}$};
%\node at (-3,7) (35) {$\bm{x_3x_5}$};
%\node at (-3,8) (36) {$\bm{x_3x_6}$};
%\node at (-4,8) (45) {$\bm{x_4x_5}$};
%\node at (-4,9) (46) {$\bm{x_4x_6}$};
%\node at (-5,10) (56) {$\bm{x_5x_6}$};

\draw[line width=0.25mm](0) --(1) --(2) -- (3) -- (4) -- (5) -- (6) -- (06);
\draw[line width=0.25mm](1)--(01) --(02) --(03) -- (04) --(05) --(06) -- (16) -- (26);% -- (36) -- (46) -- (56);
\draw[line width=0.25mm](2)--(02) --(12) --(13) -- (14) -- (15) -- (16);
\draw[line width=0.25mm](3)--(03) --(13) --(23) -- (24) -- (25) -- (26);
\draw[line width=0.25mm](4)--(04) --(14) --(24) -- (34);% -- (35) -- (36);
\draw[line width=0.25mm](5)--(05) --(15) --(25);% -- (35) -- (45) -- (46);

\draw[line width=0.75mm, dotted,  blue] (0.5+1,-0.5-1.8) -- (0.5+1,4) -- (-1.5,7) -- (-1.5,1.75) -- (-0.4,0.8) -- (-0.4,-0.3)--(-0.4+1,-0.3-1) --(0.6,-2.3)-- (0.5+1,-0.5-1.8);

\draw[line width=0.75mm, dotted,  red] (0.4+1,-0.4-1.8) -- (1.4,4) -- (0.2,5.2) -- (-1.3,5.2) -- (-2.5,4) -- (-1.3,3) -- (-1.3,1.8) --(-0.3,0.9) -- (-0.3,-0.2)-- (-0.3+1,-0.2-1) -- (0.7,-2.2) --  (0.4+1,-0.4-1.8);

\draw[line width=0.75mm, dotted,  green] (0.6+1,-0.6-1.8) -- (1.6,4.2) -- (1.6-1,4.2) --(0.6,4)-- (-1,5.3) -- (-2.5,5.3) --(-2.5,3.8) -- (-1.6,3) --(-1.6,1.6) -- (-0.5,0.7) -- (-0.5,-0.3)--(0.5,-0.3-1) --(0.5,-0.6-1.8)-- (0.6+1,-0.6-1.8);
;

\draw[line width=0.75mm, dotted,  brown] (0.7+1,-0.7-1.8) -- (0.7+1,5.3) -- (-0.5,5.3) -- (-0.5,4) -- (-2,5.5) -- (-2.7,5.5)-- (-2.7,3.8) -- (-1.8,3) --(-1.8,1.6) -- (-0.6,0.6) --  (-0.6,-0.4)--(0.4,-1.4)-- (0.4,-0.7-1.8) -- (0.7+1,-0.7-1.8);

\draw[line width=0.75mm, dotted,  yellow] (0.8+1,-0.8-1.8) -- (0.8+1,4.3)  -- (0.8-1.2,4.3) -- (0.8-1.2,4) -- (-2.6,6.2) -- (-3.5,6.2)--(-2.9,5.6)-- (-2.9,3.8) -- (-2,3) --(-2,1.6) -- (-0.7,0.5) --  (-0.6-0.1,-0.4)--(0.4-0.1,-1.4)-- (0.4-0.1,-0.7-1.9) -- (0.7+1+0.1,-0.7-1.9);

\end{tikzpicture}
}
    \caption{Five codes of dimension 19 for $m=7$. Delimited areas can be compared to different metro/bus lines, where each line has exactly 18 stops. }
    \label{fig:five_codes_K19}
\end{figure}
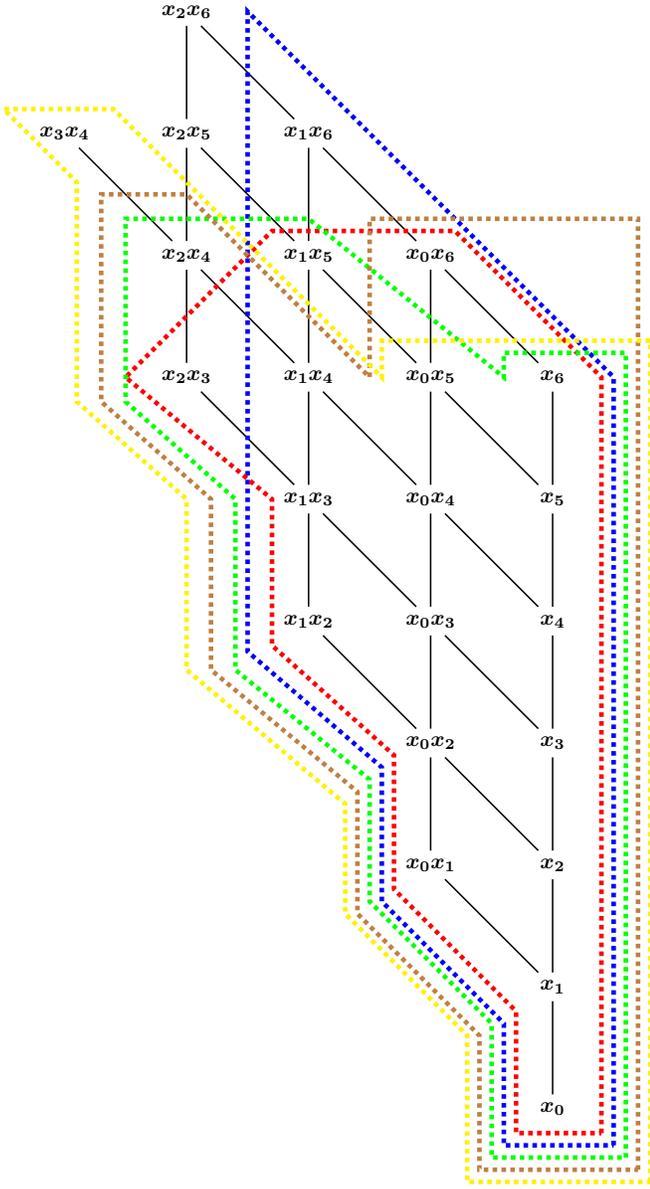

\begin{example}
    Let $m=7$ and the monomials in $\I$ are defined by $f\preceq x_3x_4,f\preceq x_2x_6$ where $|\I|=24$. We are interested in selecting 19 monomials from $\I$ to construct a subcode of dimension 19 that provides a good weight distribution. We consider five options (all being decreasing monomial codes) illustrated in Fig. \ref{fig:five_codes_K19} circled by different colors. Note that the monomial $1$ is not included in the figure. Also, all these 5 codes have a common subcode of dimension $16$ (the monomials covered by the intersection of all the lines), which is given by $f\preceq x_1x_4,f \preceq x_0x_5,f \preceq x_6.$ Let us denote this common subset of monomials by $\J.$ 

\begin{itemize}
    \item Blue lines: $\I_2=\{f\mid f\preceq x_1x_6$ and the code is $\C(\J^{*})$ where $\J^{*}=\J\cup\{x_1x_5,x_0x_6,x_1x_6\}).$ Compared to $\J$ we have added monomials $f$ with $|\lambda_f|\in\{5,5,6\}.$ The weight polynomial for this code is 
\[W(\J^{*},X)=1+ 748X^{32}+ 29760X^{48}+ 463270X^{64}. \]
    \item Yellow lines: $\I_2=\{f\mid f\preceq x_3x_4$ and the set $\J^{*}=\J\cup\{x_2x_3,x_2x_4,x_3x_4\}).$ Compared to $\J$ we have added monomials $f$ with $|\lambda_f|\in\{4,5,6\}.$ The weight polynomial for this code is 
\[W(\J^{*},X)=1+ 684X^{32}+ 22848X^{48}+ 28672x^{56}+ 419878X^{64}. \]  
    \item Red lines: $\I_2=\{f\mid f\preceq x_0x_6,f\preceq x_1x_5, f\preceq x_2x_3\}.$ Compared to $\J$ we have added $f\in\{x_2x_3,x_1x_5,x_0x_6\}$ which makes $|\lambda_f|\in\{4,5,5\}.$ Using our formula we obtain    
\[W(\J^{*},X)=1+ 556X^{32}+ 21312X^{48} + 36864X^{56}+ 406822X^{64}.\]
    \item Green lines: This code is similar to the previous one, instead of $x_0x_6$ we have $x_2x_4$ and thus we obtain the same weight distribution.
    \item Brown lines: Here we are in a similar case as the previous code, and obtain the same weight distribution,
% I:=[1,x1x7] U [1,x3x5]
%
% [ 556, 21312, 36864, 0, 406822 ]
% -----------------------------------------------------------------------------------
% S:={{1,2},{1,3},{1,4},{1,5},{1,6},{2,3},{2,4},{2,5},{3,4},{3,5},{4,5}};
   % \item Yellow lines: I:=[1,x1x6] U [1,x4x5]
% 
%
%
\end{itemize}    
\end{example}

    \midsepremove    
\begin{table}[!ht]
    \centering
    \begin{tabular}{c|c|c|c|c}
    \toprule
          $m=5$ & $m=6$ & $m=7$ & $m=8$ & $m=9$  \\
         \midrule
            \rowcolor{c1}0, 1&0, 1&0, 1&0, 1&0, 1\\\midrule
            \rowcolor{c2}0, 2&0, 2&0, 2&0, 2&0, 2\\\midrule
            \rowcolor{c3}1, 2&1, 2&1, 2&1, 2&1, 2\\
            \rowcolor{c3}0, 3&0, 3&0, 3&0, 3&0, 3\\\midrule
            \rowcolor{c4}1, 3&1, 3&1, 3&1, 3&1, 3\\
            \rowcolor{c4}0, 4&0, 4&0, 4&0, 4&0, 4\\\midrule
            \rowcolor{c5}2, 3&2, 3&2, 3&2, 3&2, 3\\
            \rowcolor{c5}1, 4&1, 4&1, 4&1, 4&1, 4\\ \cmidrule{1-1}\rowcolor{c5}\cellcolor{c6}2, 4&0, 5&0, 5&0, 5&0, 5\\\cmidrule{2-5}
            \rowcolor{c6}\cellcolor{c7}3, 4&2, 4&2, 4&2, 4&2, 4\\
            \rowcolor{c6}\cellcolor{white}&1, 5&1, 5&1, 5&1, 5\\\cmidrule{2-2}
            \rowcolor{c6}\cellcolor{white}&\cellcolor{c7}3, 4&0, 6&0, 6&0, 6\\\cmidrule{3-5}
            \rowcolor{c7}\cellcolor{white}&2, 5&3, 4&3, 4&3, 4\\\cmidrule{2-2}
            \rowcolor{c7}\cellcolor{white}&\cellcolor{c8}3, 5&2, 5&2, 5&2, 5\\\cmidrule{2-2}
            \rowcolor{c7}\cellcolor{white}&\cellcolor{c9}4, 5&1, 6&1, 6&1, 6\\\cmidrule{3-3}
            \rowcolor{c7}\cellcolor{white}&\cellcolor{white} &\cellcolor{c8} 3, 5&0, 7&0, 7\\\cmidrule{4-5}
            \rowcolor{c8}\cellcolor{white}&\cellcolor{white} & 2, 6&3, 5&3, 5\\\cmidrule{3-3}
            \rowcolor{c8}\cellcolor{white}&\cellcolor{white} &\cellcolor{c9} 4, 5&2, 6&2, 6\\
            \rowcolor{c8}\cellcolor{white}&\cellcolor{white} & \cellcolor{c9}3, 6&1, 7&1, 7\\\cmidrule{3-4}
            & & \cellcolor{c10}4, 6&\cellcolor{c9}4, 5&\cellcolor{c8}0, 8\\\cmidrule{3-3}\cmidrule{5-5}
            & & \cellcolor{c11}5, 6&\cellcolor{c9}3, 6&\cellcolor{c9}4, 5\\
            & & &\cellcolor{c9}2, 7&\cellcolor{c9}3, 6\\\cmidrule{4-4}
            & & &\cellcolor{c10}4, 6&\cellcolor{c9}2, 7\\
            & & &\cellcolor{c10}3, 7&\cellcolor{c9}1, 8\\\cmidrule{4-5}
            & & &\cellcolor{c11}5, 6&\cellcolor{c10}4, 6\\
            & & &\cellcolor{c11}4, 7&\cellcolor{c10}3, 7\\
            & & &\cellcolor{c12}5, 7&\cellcolor{c10}2, 8\\\cmidrule{4-5}
            & & &\cellcolor{c13}6, 7&\cellcolor{c11}5, 6\\
            & & &&\cellcolor{c11}4, 7\\
            & & &&\cellcolor{c11}3, 8\\
            & & &&\cellcolor{c12}5, 7\\
            & & &&\cellcolor{c12}4, 8\\
            & & &&\cellcolor{c13}6, 7\\
            & & &&\cellcolor{c13}5, 8\\
            & & &&\cellcolor{c14}6, 8\\
            & & &&\cellcolor{c15}7, 8\\
            \bottomrule
    \end{tabular}
    \caption{Ordered maximum degree monomials (degree 2, represented by indices of their variables) with respect to weight contribution (highlighted by varying colors) and reliability by PW. The monomial with indices 0,1 is the most reliable one.}
    \label{tab:weight-PW}
\end{table}

\begin{table*}[!ht]
    \centering
    \begin{tabular}{c|l}
         \toprule
         Dimension & Information set and code properties \\
         \midrule
           & $\I_2=\{x_0x_1\}$\\
         9 & $W(\I,X)=1+ 4X^{32}+ 502X^{64}+ 4X^{96}+ X^{128}$\\
           & $|\Aut(\C(\I))|=2^{33} \times 3^3 \times 5 \times 7 \times 31$\\
           \midrule
           & $\I_2=\{x_0x_1,x_0x_2\}$\\
         10 & $W(\I,X)=1+ 12X^{32}+ 998X^{64}+ 12X^{96}+ X^{128}$\\
           & $|\Aut(\C(\I))|=2^{37} \times 3^4 \times 5 \times 7 $\\
           \midrule
         \multirow{6}{*}{11}  & $\I_2=\{x_0x_1,x_0x_2,x_0x_3\}$\\
          & $W(\I,X)=1+ 28X^{32}+ 1990X^{64}+ 28X^{96}+ X^{128}$\\
           & $|\Aut(\C(\I))|=2^{40} \times 3^3 \times 7^3 $\\
           \cline{2-2}
            & $\I_2=\{x_0x_1,x_0x_2,x_1x_2\}$\\
            & $W(\I,X)=1+ 28X^{32}+ 1990X^{64}+ 28X^{96}+ X^{128}$\\
           & $|\Aut(\C(\I))|=2^{41} \times 3^4 \times 5^2 \times 7^2 $\\
           \midrule
           & $\I_2=\{x_0x_1,x_0x_2,x_1x_2,x_0x_3\}$\\
         12 & $W(\I,X)=1+ 44X^{32}+ 64X^{48}+3878X^{64}+ 64X^{80}+44X^{96}+ X^{128}$\\
           & $|\Aut(\C(\I))|=2^{43} \times 3^3 \times 7 $\\
           \midrule
           \multirow{6}{*}{13}  & $\I_2=\{x_0x_1,x_0x_2,x_0x_3,x_1x_2,x_0x_4\}$\\
          & $W(\I,X)=1+ 76X^{32}+ 192X^{48}+7654X^{64}+ 192X^{80}+76X^{96}+ X^{128}$\\
           & $|\Aut(\C(\I))|=2^{44} \times 3^5$\\
           \cline{2-2}
            & $\I_2=\{x_0x_1,x_0x_2,x_0x_3,x_1x_2,x_1x_3\}$\\
            & $W(\I,X)=1+ 76X^{32}+ 192X^{48}+7654X^{64}+ 192X^{80}+76X^{96}+ X^{128}$\\
           & $|\Aut(\C(\I))|=2^{45} \times 3^3 \times 7 $\\
           \midrule
             & $\I_2=\{x_0x_1,x_0x_2,x_0x_3,x_1x_2,x_1x_3,x_0x_4\}$\\
            14& $W(\I,X)=1+ 108X^{32}+ 576X^{48}+15014X^{64}+ 576X^{80}+96X^{96}+ X^{128}$\\
           & $|\Aut(\C(\I))|=2^{45} \times 3^2 $\\
           \midrule
             \multirow{9}{*}{15}  & $\I_2=\{x_0x_1,x_0x_2,x_0x_3,x_1x_2,x_1x_3,x_0x_4,x_0x_5\}$\\
            & $W(\I,X)=1+ 172X^{32}+ 1344X^{48}+29734X^{64}+ 1344X^{80}+176X^{96}+ X^{128}$\\
           & $|\Aut(\C(\I))|=2^{44} \times 3^3 $\\
           \cline{2-2}
           & $\I_2=\{x_0x_1,x_0x_2,x_0x_3,x_1x_2,x_1x_3,x_0x_4,x_1x_4\}$\\
            & $W(\I,X)=1+ 172X^{32}+ 1344X^{48}+29734X^{64}+ 1344X^{80}+176X^{96}+ X^{128}$\\
           & $|\Aut(\C(\I))|=2^{45} \times 3^3 \times 7$\\
           \cline{2-2}
           & $\I_2=\{x_0x_1,x_0x_2,x_0x_3,x_1x_2,x_1x_3,x_0x_4,x_2x_3\}$\\
            & $W(\I,X)=1+ 172X^{32}+ 1344X^{48}+29734X^{64}+ 1344X^{80}+176X^{96}+ X^{128}$\\
           & $|\Aut(\C(\I))|=2^{45} \times 3^2\times 7 $\\
           \midrule
              \multirow{9}{*}{16}  & $\I_2=\{x_0x_1,x_0x_2,x_0x_3,x_1x_2,x_1x_3,x_0x_4,x_0x_5,x_2x_3\}$\\
            & $W(\I,X)=1+ 236X^{32}+ 3136X^{48}+58790X^{64}+ 3136X^{80}+236X^{96}+ X^{128}$\\
           & $|\Aut(\C(\I))|=2^{43} \times 3^3 \times 7$\\
           \cline{2-2}
           & $\I_2=\{x_0x_1,x_0x_2,x_0x_3,x_1x_2,x_1x_3,x_0x_4,x_1x_4,x_0x_5\}$\\
            & $W(\I,X)=1+ 236X^{32}+ 3136X^{48}+58790X^{64}+ 3136X^{80}+236X^{96}+ X^{128}$\\
           & $|\Aut(\C(\I))|=2^{43} \times 3 \times 7$\\         
           \cline{2-2}
           & $\I_2=\{x_0x_1,x_0x_2,x_0x_3,x_1x_2,x_1x_3,x_0x_4,x_2x_3,x_1x_4\}$\\
            & $W(\I,X)=1+ 236X^{32}+ 3136X^{48}+58790X^{64}+ 3136X^{80}+236X^{96}+ X^{128}$\\
           & $|\Aut(\C(\I))|=2^{45} \times 3^3 $\\         
           \bottomrule
    \end{tabular}
    \caption{Weight Distribution and permutation group of the codes constructed with our algorithm, based on weight contribution and reliability, for $m=7$ and dimensions between $9$ and $16.$}
    \label{tab:code-properties-m7}
\end{table*}

%\subsection{Beta-expansion and weight contribution}
In \cite{he}, the PW order relation was defined as follows: \[f\leq_{\beta}g\text{ if }\sum_{i\in\ind{f}} \beta^{i}\leq \sum_{i\in\ind{g}} \beta^{i}.\] 

Let us consider the proposed value $\beta=2^{\frac{1}{4}}$. We notice that for $m<7$ the order induced by beta-expansion respects $\preccurlyeq_{\wm}.$ However, starting from $m=7$ some pair of monomials monomials will be in conflict with the two order relations, e.g., $x_2x_5\leq_{\beta} x_0x_6$ while $x_0x_6\preccurlyeq_{\wm}x_2x_5.$ The number of conflicting pairs increases with $m.$ Hence, we need to define a rule which allows the orders relations to blend. We propose to order first w.r.t. $\preccurlyeq_{\wm}$ and then use $\leq_{\beta}$ on the non-comparable elements w.r.t. $\preccurlyeq_{\wm}.$ 
%For $m=5$ we have the following ordered monomials 
%\[x_0x_1, x_0x_2, x_1x_2, x_0x_3, x_1x_3,x_0x_4, x_2x_3,x_1x_4,x_2x_4,x_3x_4.\]
In Table \ref{tab:weight-PW}, we illustrate this new order relation for $5\leq m\leq 9$, where the monomials of maximum degree are listed. The level of weight contribution of monomials can be visualized as a color spectrum, where darker shades correspond to higher contributions from the monomial in the weight distribution. 

\subsection{Permutation group}
The order $\preccurlyeq_{\wm}$ induces new symmetries which generate properties yet to be discovered. For example, elements at the same level $S_l$ have identical weight contribution, thus defining different codes with the same weight distribution. However, the codes constructed based on our algorithm are not permutation equivalent. While two permutation equivalent codes will have an identical weight distribution, the converse is not always true, and few results are known in this direction \cite{cheon2006}. The following examples further clarify the matter.
\begin{example}
    Let us consider the codes in Example \ref{ex:5}. The three codes have different permutation groups, and thus are not equivalent. $\C(\I\setminus \{x_0x_5\})$ has a group of order $2^{30}\times 3^2$, $\C(\I\setminus \{x_1x_4\})$ of order $2^{28}\times 3^3\times 7$, while $\C(\I\setminus \{x_2x_3\})$ of order $2^{28}\times 3\times 7.$ 
\end{example}
\begin{example}
For $m=7$ and the code dimensions between $9$ and $16$,  we find the corresponding codes and their properties based on the poset of weight contribution and reliability, as illustrated in Table \ref{tab:code-properties-m7}. The weight distribution is computed with our formula and the permutation group ($\Aut(\C(\I))$) is obtained using MAGMA's permutation group function. The order of the permutation group of $\R(2,7)$ is $2^{28} \times 3^4 \times 5 \times 7^2 \times 31 \times 127.$ In general, our codes have a smaller permutation group than $\R(2,7).$ However, the code of dimension 11, defined by $\I_2=\{x_0x_1,x_0x_2,x_1x_2\}$ has a permutation group about 10 times larger than $\R(2,7)$. Also, codes with same dimension have identical weight distribution. 
\end{example}

%\clearpage
\printbibliography
\clearpage
\appendices

\section{Technical facts}
\begin{lemma}[Proposition 3.7.12~\cite{dragoi17thesis}]\label{lem:prod-lin-form-distinct-var}
    Let $P=\prod_{j=1}^{l}y_j$ be a product of $l$ independent linear forms $y_j$ each having maximum variables $x_{i_j}$ (with respect to $\preceq$). Then $P$ can be written as $P=\prod_{j=1}^{l}y_j^{*}$ where all maximum variables $x_{i_j^{*}}$ in $y_j$ are pairwise distinct.
\end{lemma}

Straightforward, notice that the total number of distinct variables in a product of $l$ independent linear forms should always be at least equal to $l.$ Also, the proof of Theorem \ref{thm:Kasami-Tokura-1} comes directly from Theorem \ref{thm:Kasami-Tokura} and Lemma \ref{lem:prod-lin-form-distinct-var}.

\begin{lemma}[\cite{rowshan1.5w}]\label{lem:decomp-ltam-gcd}Let $f,g\in\I_r$ and $h=\gcd(f,g)\in\Mon.$ Then 
\begin{multline*}
    \Alow\cdot h\cdot \left(\Alow\cdot \frac{f}{h}+\Alow\cdot \frac{g}{h}\right)\\=
    \Alow_h\cdot h\cdot \left(\Alow_{f}\cdot \frac{f}{h}+\Alow_g\cdot \frac{g}{h}\right).
\end{multline*}
\end{lemma}

Next result is a generalisation of Lemma \ref{lem:decomp-ltam-gcd} to a finite sequence of monomials. 
\begin{lemma}
    Let $l$ be a fixed positive integer and let $f_i\in\I_r$ for $i\in[1,l]$ be a finite sequence of monomials. Let $h=\gcd(f_i,f_j)\in\Mon$ for all $i,j\in[1,l], i\neq j.$ Then 
\begin{multline*}
    \Alow\cdot h\cdot \sum\limits_{i=1}^l\Alow\cdot \frac{f_i}{h}\\=
    \Alow_h\cdot h\cdot \sum\limits_{i=1}^l\Alow_{f_i}\cdot \frac{f_i}{h}.
\end{multline*}
\end{lemma}

The proof of this lemma is rather straightforward using Lemma \ref{lem:prod-lin-form-distinct-var}.
\section{Proof Of Theorem \ref{thm:equality-orbits-typeII}}

\begin{IEEEproof}
 By Theorem \ref{thm:Kasami-Tokura}, any codeword $\ev(P)$ satisfying the required weight condition and being of Type II is the evaluation of a polynomial $P=y_1\dots y_{r-2}(y_{r-1}y_{r}+\dots+y_{r+2\mu-3}y_{r+2\mu-2})$ where $m-r+2\geq 2\mu\geq 2.$

    Consider a codeword $\ev(P)$ with $P=y_1\dots y_{r-2}(y_{r-1}y_{r}+\dots+y_{r+2\mu-3}y_{r+2\mu-2}).$ We can apply Lemma \ref{lem:prod-lin-form-distinct-var} and recollect the maximum variables in $P_1=y_1\dots y_{r-2}y_{r-1}y_{r}$ to $ P_{\mu}=y_1\dots y_{r-2}y_{r+2\mu-3}y_{r+2\mu-2}$ in order to construct $h=x_{i_1}\dots x_{i_{r-2}}$ and $f_1/h=x_{i_{r-1}}x_{i_{r}},\dots, f_{\mu}/h=x_{i_{r+2\mu-3}}x_{i_{r+2\mu-2}}.$ Notice that while $i_1,\dots i_{r-2}$ are all pairwise distinct, the indices $i_{r-1},i_{r},\dots,i_{r+2\mu-2}$ do not have to be pairwise distinct. By Theorem \ref{thm:1.5d} any pair ($f_i/h,f_j/h)$ can be transformed into a new pair ($f_i^*/h,f_j^*/h)$ satisfying $\gcd(f_i^{*}/h,f_j^{*}/h)=1$. By ordering the monomials $f_1/h,f_2/h,\dots ,f_{\mu}/h$ (with respect to the decreasing index order on the support $\ind(f_i/h)$) and applying Theorem \ref{thm:1.5d} we can transform all monomials $f_j/h$ into new monomials $f_j^{*}/h$ such that they are pairwise co-prime. Since $P\in \mathrm{span}(\I)$ we either have all monomials $f_i^*\in \I$ and then the proof is finished, or there might be pair of monomials $(f_i^*,f_j^*)$ that are not in $\I.$ However, since $P\in\mathrm{span}(\I)$ this implies that $f_i^*=f_j^*$ which is impossible since $\gcd(f_i^*,f_j^*)=1.$  
 \end{IEEEproof}

From Theorem \ref{thm:equality-orbits-typeII} we deduce
\begin{corollary}\label{cor:typeII}
    Let $\I$ be a decreasing monomial set with $r=\max_{f\in\I}\deg(f).$ Let $\w_{\mu}=2^{m+1-r}-2^{m+1-r-\mu}$ with $2\leq 2\mu \leq m-r+2.$ Then the set of all $\w_{\mu}$-weight codewords of type II is 
    \begin{multline}
        W_{\w_{\mu}}=\\ 
        \bigcup\limits_{\substack{i\in[1,\mu], f_i\in\I_r\\ \forall\; i,j \in [1,\mu], i\neq j, \; h=\gcd(f_i,f_j)\\h \in \I_{r-2}}}\Alow_h\cdot h\cdot\sum_{i=1}^{\mu}\Alow_{f_i}\cdot \frac{f_i}{h}
    \end{multline}      
\end{corollary}

%\section{Proof of Proposition \ref{pr:degree-two-minkowski-mu}}
\section{Proof of Proposition \ref{pr:cardinal-product-orbits-type2}}
Let us first define and recall the nescessary ingredients for our proof.
\begin{definition}Let $P,P^{*}\in\Alow\cdot f$ and $Q,Q^{*}\in\Alow\cdot g.$ we say that 
    \begin{itemize}
        \item $(P,Q),(P^{*},Q^{*})$ produces a collision for addition if $P+Q=P^{*}+Q^{*}$ with $P\neq P^{*}, Q\neq Q^{*}.$
        \item $(P,Q),(P^{*},Q^{*})$ produces a collision for multiplication if $PQ=PQ^{*}$ or $PQ=P^{*}Q^{*}$ or $PQ=P^{*}Q$ with $P\neq P^{*}, Q\neq Q^{*}.$
    \end{itemize}
\end{definition}

Let us provide an example.
\begin{example}
   Let $f=x_2x_6,g=x_3x_5$ and $P=(x_2+x_1)(x_6+x_4+1)\in \Alow\cdot f$ and $Q=(x_3+x_2+x_0+1)(x_5+x_4+x_2)\in\Alow\cdot g$. We can create two non-trivial distinct pairs $P^{*},Q^{*}$ 
    \begin{itemize}
        \item $P^{*}=(x_2+x_1)(x_6+x_4+x_3+x_2+x_0)$ and $Q^{*}=(x_3+x_2+x_0+1)(x_5+x_4+x_1)$ 
        \item $P^{*}=(x_2+x_1)(x_6+{\color{blue}x_5}+x_3+x_2+x_1+x_0)$ and $Q^{*}=(x_3+x_1+x_0+1)(x_5+x_4+x_1)$
    \end{itemize}
 Let $f=x_2x_4,g=x_0x_3.$ We obtain that $f+g=x_2(x_4+x_0)+x_0(x_3+x_2).$  
\end{example}

Recall from \cite{rowshan1.5w} that we have   
\begin{proposition}\label{pr:degree-two-minkowski}
     Let $\I\subseteq\Mon$ be a decreasing monomial set and $f=x_{i_1}x_{i_2}$ and $g=x_{j_1}x_{j_2}$ with $\gcd(f,g)=1.$ Then 
\begin{multline}
%\begin{equation}
    |\Alow\cdot f+\Alow\cdot g|=\\ \frac{|\Alow\cdot f|\times |\Alow\cdot g|}{2^{\alpha_{{f},{g}}}}.
%\end{equation}
\end{multline}
\end{proposition}

We will first demonstrate the following result.

\begin{proposition}\label{pr:degree-two-minkowski-mu}
     Let $\I\subseteq\Mon$ be a decreasing monomial set and $\mu\geq 2$ a positive integer. Also, let $f_i\in \I_2$ with $\gcd(f_i,f_j)=1$ for any pair $(i,j)\in[1,\mu]\times [1,\mu]$ with $i\neq j.$ Then 
\begin{multline}
%\begin{equation}
    \left|\sum\limits_{i=1}^{\mu}\Alow\cdot f_i\right|= \frac{\prod\limits_{i=1}^{\mu}|\Alow\cdot f_i|}{2^{\sum\limits_{i\neq j}\alpha_{{f_i},{f_j}}}}.
%\end{equation}
\end{multline}
\end{proposition}

\begin{IEEEproof}
    We will show that there are no other possible collisions for addition, except for those generated by pairs $(f_i,f_j).$ Therefore, we will suppose no collisions of pairs $(f_i,f_j)$ are allowed and demonstrate that any tuple $(f_1,\dots, f_{\mu})$ does not admit any other collision. 
    
    Let $f_1=x_{i_1}x_{i_2},\dots,f_{\mu}=x_{i_{2\mu-1}}x_{i_{2\mu}}$ with $i_{2j-1}>i_{2j}$ for all $j\in[1,\mu].$ Now let $P_j=(\bB_j,\varepsilon_j)\cdot f_j$ and $P_j^{*}=(\bB_j^{*},\varepsilon_j^{*})\cdot f_j$ be polynomials from the orbits $\Alow\cdot f_j$ and suppose there is a collision $\sum \limits_{j=1}^{\mu}P_j =  \sum \limits_{j=1}^{\mu}P_j^{*}.$ We then have
    \begin{multline}\label{eq:pr2-1}
         \sum\limits_{j=1}^{\mu}(x_{i_{2j-1}}+l_{i_{2j-1}})(x_{i_{2j}}+l_{i_{2j}})
         \\=\sum\limits_{j=1}^{\mu}(x_{i_{2j-1}}+l_{i_{2j-1}}^{*})(x_{i_{2j}}+l_{i_{2j}}^{*})
     \end{multline}
      and 
      \begin{equation}\label{eq:pr2-2}
          \forall\; i,j\in [1,\mu], i\neq j, \text{ we have } P_i + P_j \neq P_i^{*} + P_j^{*}.
      \end{equation}

    Equation \eqref{eq:pr2-2} translates the no collision condition on all pairs $(f_i,f_j)$, which is equivalent to $\alpha_{f_i,f_j}=0.$ For example $\alpha_{f_1,f_2}=0$ implies $i_1>i_2>i_3>i_4$ or $i_3>i_4>i_1>i_2.$ Without loss of generality we can consider $i_1>i_2>\dots > i_{2\mu-1}>i_{2\mu}.$   
     Extracting the coefficient of the maximum variable ($x_{i_1}$) from \eqref{eq:pr2-1} we obtain \begin{equation}
         x_{i_2}+l_{i_2}=x_{i_2}+l_{i_2}^{*} 
     \end{equation}
     which implies 
    \begin{multline}\label{eq:pr2-3}
        (l_{i_1}+l_{i_1}^{*})(x_{i_2}+l_{i_2}) =
        \sum\limits_{j=2}^{\mu}(x_{i_{2j-1}}+l_{i_{2j-1}})(x_{i_{2j}}+l_{i_{2j}})\\
        +\sum\limits_{j=2}^{\mu}(x_{i_{2j-1}}+l_{i_{2j-1}}^{*})(x_{i_{2j}}+l_{i_{2j}}^{*}).
    \end{multline}
    Let $j_1$ denote the index of the maximum variable in $l_{i_1}+l_{i_1}^{*}.$ By definition of $\Alow_f\cdot f$ we have that $j_1$ is different from $i_2$ and $j_1<i_1.$\\
    Suppose $j_1>i_2$, and since $i_2$ is strictly greater than all indices $i_j$ with $j\in[3,2\mu]$, we can extract the coefficient of $x_{j_1}$ from \eqref{eq:pr2-3} to obtain 
    \begin{equation}
    x_{i+2}+l_{i_2}=0,    
    \end{equation}
     which is impossible. \\
     Therefore, $j_1<i_2.$ Extracting the coefficient of $x_{i_2}$ from \eqref{eq:pr2-3} we obtain
     \begin{equation}
         l_{i_1}=l_{i_1}^{*},
     \end{equation}
    which is impossible.
    In conclusion, all collision are solely generated by collisions between pairs of monomials. Hence, the cardinality of the set $\sum\limits_{i=1}^{\mu}\Alow\cdot f_i$ equals $\prod\limits_{i=1}^{\mu}\left|\Alow\cdot f_i\right|$ divided by the total number of collisions, which is $2^{\sum\limits_{i,j\in[1,\mu]; i\neq j}\alpha_{f_i,f_j}}$
\end{IEEEproof}

Now we can proceed to the proof of Proposition \ref{pr:cardinal-product-orbits-type2}.
\begin{IEEEproof}
       We need to prove that there are no multiplicative collisions and then simply apply Proposition \ref{pr:no-colision-orbits-typeII}. So, let us suppose there are polynomials $H,H^{*}\in\Alow_h\cdot h$ and $P,P^{*}\in \sum_{i\in[1,\mu]}\Alow_{f_i}\cdot \frac{f_i}{h}$ s.t. $HP=H^{*}P^{*}.$ Since $h$ is a product of variables that are not present in $P$ or $P^{*}$ extracting the coefficient of $h$ from $HP$ and $H^{*}P^{*}$ implies $P=P^{*}.$ This implies $P(H+H^{*})=0.$ By definition we can set $H=\prod_{j\in [1,r-2]}(x_{i_j}+l_{i_j})$, $H^{*}=\prod_{j\in [1,r-2]}(x_{i_j}+l_{i_j}^{*}).$  
Hence, we obtain 
\begin{equation}\label{eq:pr3-1}
P\left(\sum\limits_{j\in[1,r-2]} \frac{h}{x_{i_j}}l_{i_j}^{\prime}
        +\dots  +\prod\limits_{j\in[1,r-2]}l_{i_j}\prod\limits_{j\in[1,r-2]}l_{i_j}^{*}\right)=0
        \end{equation}
where $l_{i_j}^{\prime}\triangleq l_{i_j}+l_{i_j}^*.$
Since all the monomials $h/x_{i_j}$ are unique in the expansion (by definition of the $\Alow$), we deduce 
\[P(l_{i_j}+l_{i_j}^{*})=0,\forall j \in [1,r-2].\]

Since $P=(x_{i_{r-1}}+l_{i_{r-1}})(x_{i_{r}}+l_{i_r})+\dots +(x_{i_{r+2\mu-3}}+l_{i_{r-+2\mu-3}})(x_{i_{r+2\mu-2}}+l_{i_{r+2\mu-2}})$ and the variables $x_{i_{r-1}},\dots x_{i_{r+2\mu-2}}$ are all distinct, the last equation can hold only if $l_{i_j}+l_{i_j}^{*}=0$, fact that ends the proof.
\end{IEEEproof}

\section{Proof of Theorem \ref{thm:formula_typeII}}
We shall first demonstrate that two distinct tuples of monomials define disjoint orbits.
\begin{proposition}\label{pr:no-colision-orbits-typeII}
    Let $\I$ be a decreasing monomial set and let $(f_i)_{i\in[1,\mu]},(f^{*}_i)_{i\in[1.\mu]}\in \I_r^{\mu}$ with $\deg(h)=\deg(h^{*})=r-2$, where $h=\gcd(f_i,f_j)$ and $h^{*}=\gcd(f_i^{*},f_j^{*})$ for all distinct $i,j\in[1,\mu].$ 
    
    Then if $\left|\{f_i\}_{i\in[1,\mu]}\bigcap \{f^{*}_i\}_{i\in[1,\mu]}\right|\leq \mu-1$ the two sets $\Alow_h\cdot h\cdot\sum_{i=1}^{\mu}\Alow_{f_i} \cdot \frac{f_i}{h}$ and $\Alow_{h^{*}}\cdot h^{*}\cdot\sum_{i=1}^{\mu}\Alow_{f^{*}_i} \cdot \frac{f^{*}_i}{h^{*}}$ are disjoint.
\end{proposition}

In order to prove our result we will recall a useful fact from \cite{rowshan1.5w}.
\begin{proposition}[\cite{rowshan1.5w}]\label{pr:no-colision-1.5dorbits}
    Let $\I$ be a decreasing monomial set and let $(f,g),(f^{*},g^{*})\in \I_r\times\I_r$ with $(f,g)\neq(f^{*},g^{*})$ and $\deg(h)=\deg(h^{*})=r-2$, where $h=\gcd(f,g)$ and $h^{*}=\gcd(f^{*},g^{*}).$ Then the sets $\Alow_h\cdot h\cdot(\Alow_{f} \cdot \frac{f}{h}+\Alow_{g}\cdot \frac{g}{h})$ and $\Alow_{h^{*}}\cdot h^{*}\cdot(\Alow_{f^{*}} \cdot \frac{f^{*}}{h^{*}}+\Alow_{g^{*}}\cdot \frac{g^{*}}{h^{*}})$ are disjoint.
\end{proposition}

We can now proceed to the proof of our result.
\begin{IEEEproof}
    We shall use induction on $\mu.$ The case $\mu=2$ is demonstrated in Proposition \ref{pr:no-colision-1.5dorbits}.  
    Let us also consider the case $\mu=3.$ In other words there are monomials $(f_1,f_2,f_3),(f_1^{*},f_2^{*},f_3^{*})\in \I_r^3$ with $h=\gcd(f_1,f_2,f_3)\in \I_{r-2}$ and $h^{*}=\gcd(f_1^{*},f_2^{*},f_3^{*})\in \I_{r-2}.$ By absurd suppose the intersection of the two orbits is non-trivial,i.e., there is a polynomial $P$ which belongs to both orbits. We can then write
    \begin{equation}\label{eq:Phh*}
        P=H(F_1+F_2+F_3)=H^{*}(F_1^{*}+F_2^{*}+F_3^{*}).
    \end{equation}where $H\in \Alow_{h}\cdot h,H^{*}\in \Alow_{h^{*}}\cdot h^{*}$ and $F_i\in \Alow_{f_i}\cdot f_i/h,F_i^{*}\in \Alow_{f_i^{*}}\cdot f_i^{*}/h^{*}.$ By definition of the $\Alow$ there are either 1, 2 or 3 maximum monomials in the two terms. We can not have 3 distinct maximum monomials since this would imply that $f_1,f_2,f_3$ are not comparable w.r.t. $\preceq.$ The same would be valid for $f_1^{*},f_2^{*},f_3^{*}$, and thus we would have $\{f_1,f_2,f_3\}=\{f_1^{*},f_2^{*},f_3^{*}\}$, which contradicts our assumption. 

    The case of two maximum variables: $f_1+f_2$ and $f_1^{*}+f_2^{*}$ ($f_1$ and $f_2$ being non-comparable) if $f_3\preceq f_2$ or $f_3\preceq f_1$ and if $f_3^{*}\preceq f_2^{*}$ or $f_3^{*}\preceq f_1^{*}.$ This implies that $f_1+f_2=f_1^{*}+f_2^{*}$ which means that $f_1=f_1^{*}$ and $f_2=f_2^{*}$ or $f_1=f_2^{*}$ and $f_2=f_1^{*}.$ Since $h=\gcd(f_1,f_2)$ and $h^{*}=\gcd(f_1^{*},f_2^{*})$ we have $h=h^{*}$. Let $F_1=(x_{i_1}+l_{i_1})(x_{i_2}+l_{i_2}), F_2=(x_{i_3}+l_{i_3})(x_{i_4}+l_{i_4}), F_3=(x_{i_5}+l_{i_5})(x_{i_6}+l_{i_6})$ and $F_1^{*}=(x_{i_1}+l_{i_1}^{*})(x_{i_2}+l_{i_2}^{*}), F_2^{*}=(x_{i_3}+l_{i_3}^{*})(x_{i_4}+l_{i_4}^{*}), F_3^{*}=(x_{i_5^{*}}+l_{i_5}^{*})(x_{i_6^{*}}+l_{i_6}^{*})$ with $i_1<i_2,i_3<i_4,i_5<i_6$ and $i_5^{*}<i_6^{*}.$ Extracting the coefficient of $h$ from both sides of \eqref{eq:Phh*} we obtain
   % \begin{equation*}
   %     F_1+F_2+F_3=F_1^{*}+F_2^{*}+F_3^{*}
   % \end{equation*}
\begin{multline}\label{eq:Phh*f}
        0=(x_{i_1}+l_{i_1})(x_{i_2}+l_{i_2})-(x_{i_1}+l_{i_1}^{*})(x_{i_2}+l_{i_2}^{*})\\
        +(x_{i_3}+l_{i_3})(x_{i_4}+l_{i_4})-(x_{i_3}+l_{i_3}^{*})(x_{i_4}+l_{i_4}^{*})\\
        +(x_{i_5}+l_{i_5})(x_{i_6}+l_{i_6})-(x_{i_5^{*}}+l_{i_5}^{*})(x_{i_6^{*}}+l_{i_6}^{*}).
    \end{multline}
    
   Suppose that $f_3\preceq f_2.$ Since $f_1,f_2$ are not comparable this means $i_3<i_1<i_2<i_4$ or $i_1<i_3<i_4<i_2$. Also, using $\gcd(f_1,f_3)=\gcd(f_2,f_3)=\gcd(f_1,f_2)=h$ we have that $i_5<i_3,i_6<i_4.$ 
   \begin{itemize}
       \item  $i_5<i_3<i_1<i_2<i_4,i_6<i_4$, then $x_{i_4}$ is the maximum variable in the set of variables in $f_1f_2f_3.$ Extracting the coefficient of $x_{i_4}$ from equation \eqref{eq:Phh*f} gives the following two cases.
   \begin{itemize}
       \item if $i_6<i_4$ and $i_6^{*}<i_4$ we have
       \begin{equation}
       x_{i_3}+l_{i_3}=x_{i_3}+l_{i_3}^{*}    
       \end{equation} 
       This means $l_{i_3}=l_{i_3}^{*}$ and 
       \begin{multline}\label{eq:Phh*f1}
        0=(x_{i_1}+l_{i_1})(x_{i_2}+l_{i_2})-(x_{i_1}+l_{i_1}^{*})(x_{i_2}+l_{i_2}^{*})\\
        +(x_{i_3}+l_{i_3})(l_{i_4}-l_{i_4}^{*})\\
        +(x_{i_5}+l_{i_5})(x_{i_6}+l_{i_6})-(x_{i_5^{*}}+l_{i_5}^{*})(x_{i_6^{*}}+l_{i_6}^{*}).
    \end{multline}
    The next maximum variable can be either $x_{i_2},x_{i_6}$ or $x_{i_6^{*}}.$ Notice that $i_2\neq i_6$ and $i_2\neq i_6^{*}.$ Let us consider each case separately.
        \begin{enumerate}
            \item $x_{i_2}.$ Extracting the coefficient of $x_{i_2}$ from \eqref{eq:Phh*f1} gives either
            \begin{equation}\label{eq:x2case1}
       x_{i_1}+l_{i_1}=x_{i_1}+l_{i_1}^{*}    
       \end{equation} 
       or 
         \begin{equation}\label{eq:x2case2}
       x_{i_1}+l_{i_1}=x_{i_1}+l_{i_1}^{*} + x_{i_3}+l_{i_3}    
       \end{equation}
       From \eqref{eq:x2case1} we deduce $l_{i_1}=l_{i_1}^{*}$ and 
       \begin{multline}\label{eq:Phh*f2}
        0=(x_{i_1}+l_{i_1})(l_{i_2}-l_{i_2}^{*})
        +(x_{i_3}+l_{i_3})(l_{i_4}-l_{i_4}^{*})\\
        +(x_{i_5}+l_{i_5})(x_{i_6}+l_{i_6})-(x_{i_5^{*}}+l_{i_5}^{*})(x_{i_6^{*}}+l_{i_6}^{*}).
    \end{multline}
    The next maximum variable could be either $x_{i_6}$ or $x_{i_1}.$ 
    \begin{enumerate}
        \item $x_{i_1}.$ If $x_{i_1}$ does not belong to $l_{i_4}-l_{i_4}^{*}$ extracting the coefficient of $x_{i_1}$ from \eqref{eq:Phh*f2} gives $l_{i_2}=l_{i_2}^{*}\Rightarrow F_2=F_2^{*}$ and thus using Proposition \ref{pr:no-colision-1.5dorbits} the proof is finished for this case. \\
        If $x_{i_1}$ belongs to $l_{i_4}-l_{i_4}^{*}$ extracting the coefficient of $x_{i_1}$ from \eqref{eq:Phh*f2} gives $l_{i_2}-l_{i_2}^{*} =x_{i_3}+l_{i_3}$
        which implies 
        \begin{multline}\label{eq:Phh*f3}
            (x_{i_3}+l_{i_3})(x_{i_1}+l_{i_1}+l_{i_4}+l_{i_4}^{*})=\\
            (x_{i_5}+l_{i_5})(x_{i_6}+l_{i_6})-(x_{i_5^{*}}+l_{i_5}^{*})(x_{i_6^{*}}+l_{i_6}^{*})
        \end{multline}
        Notice that the maximum variables in $x_{i_1}+l_{i_1}+l_{i_4}+l_{i_4}^{*}$ is strictly smaller than $x_{i_1}.$If $i_6<i_3$ \eqref{eq:Phh*f3} can not hold (see Proposition \ref{pr:degree-two-minkowski}). So $i_3<i_6<i_1.$ If $x_{i_1}+l_{i_1}+l_{i_4}+l_{i_4}^{*}$ does not contain $x_{i_6}$ then we need to have $i_6=i_6^{*}$ and \begin{equation}
            x_{i_5}+l_{i_5}=x_{i_5^{*}}+l_{i_5}^{*}.
        \end{equation}
        This implies $f_3=f_3^{*}$ which is impossible.
        \item The same arguments hold for $x_{i_6}.$  
    \end{enumerate}
    From \eqref{eq:x2case2} we have 
         \begin{multline}\label{eq:Phh*f22}
        0=(x_{i_1}+l_{i_1})(l_{i_2}+l_{i_2}^{*})-(x_{i_3}+l_{i_3})(x_{i_2}+l_{i_2}^{*}+l_{i_4}+l_{i_4}^{*})\\
        +(x_{i_5}+l_{i_5})(x_{i_6}+l_{i_6})-(x_{i_5^{*}}+l_{i_5}^{*})(x_{i_6^{*}}+l_{i_6}^{*}).
    \end{multline}
    Notice that the maximum variable in $x_{i_2}+l_{i_2}^{*}+l_{i_4}+l_{i_4}^{*}$ is strictly smaller than $x_{i_2}.$ Notice that the last equation resumes to \eqref{eq:Phh*f2} and hence the proof is identical for this case. 
    \item $x_{i_6}$ or $x_{i_6^{*}}.$ Here we can have the following cases 
    \begin{equation}\label{eq:casex6}
        i_6=i_6^{*}\Rightarrow x_{i_5}+l_{i_5}=x_{i_5^{*}}+l_{i_5}^{*}
    \end{equation}
     \begin{equation*}
        x_{i_5}+l_{i_5}=x_{i_3}+l_{i_3}\quad (\text{impossible})
    \end{equation*}
    \begin{equation*}
        x_{i_5^{*}}+l_{i_5}^{*}=x_{i_3}+l_{i_3}\quad (\text{impossible})
    \end{equation*}
\begin{equation*}
    x_{i_5}+l_{i_5}+x_{i_5^{*}}+l_{i_5}^{*}=x_{i_3}+l_{i_3}\quad (\text{impossible})
    \end{equation*}
    Hence, solely \eqref{eq:casex6} can hold. But this leads to $f_3=f_3^{*}$ which is impossible and ends the proof for this sub-case.
        \end{enumerate}
       \item if $i_6>i_4$ the same arguments hold.
   \end{itemize}
   \item The arguments for the case $i_1<i_3<i_4<i_2$ $i_5<i_3,i_6<i_4$ are similar and thus omitted.
   \end{itemize}
    Remains the case where we have one maximum monomial, say $f_1.$ This implies that $f_1=f_1^{*}.$ If $h=h^{*}$ we can apply the same arguments as in the previous case. If not, $h\neq h^{*}$, then $\deg(\gcd(h,h^{*}))\leq 2$ because $\deg(f/h)=\deg(f^{*}/h^{*})=2.$ Then we can put $h=x_{i_1}\dots x_{i_{r-4}}x_{i_{r-3}}x_{i_{r-2}},h^{*}=x_{i_1}\dots x_{i_{r-4}}x_{i_{r-3}^{*}}x_{i_{r-2}^{*}}$ and $f=x_{i_1}\dots x_{i_{r-4}}x_{i_{r-3}}x_{i_{r-2}}x_{i_{r-3}^{*}}x_{i_{r-2}^{*}}.$ Taking the coefficient of $h/\gcd(h,h^{*})$ in \eqref{eq:Phh*} gives 
    \begin{multline}\label{eq:Phffx3}
        (x_{i_{r-3}}+l_{i_{i-3}})(x_{i_{r-2}}+l_{i_{i-2}})\left(F_1+F_2+F_3\right)=\\
        (x_{i_{r-3}^{*}}+l_{i_{i-3}^{*}}^{*})(x_{i_{r-2}^{*}}+l_{i_{i-2}^{*}}^{*})\left(F_1^{*}+F_2^{*}+F_3^{*}\right)
    \end{multline}
        Suppose the maximum variable in $x_{i_{r-3}}x_{i_{r-2}}x_{i_{r-3}^{*}}x_{i_{r-2}^{*}}$ is $x_{i_{r-3}}.$ Extracting the coefficient of $x_{i_{r-3}}$ from \eqref{eq:Phffx3} gives the following.
         \begin{enumerate}
             \item If $x_{i_{r-3}}$ is strictly greater than the variables of $f_2^{*}/h^{*},f_3^{*}/h^{*}$  
    \begin{multline}\label{eq:Phffx3x2}
        (x_{i_{r-2}}+l_{i_{i-2}})\left(F_1+F_2+F_3\right)\\=
        (x_{i_{r-3}^{*}}+l_{i_{i-3}^{*}}^{*})(x_{i_{r-2}^{*}}+l_{i_{i-2}^{*}}^{*})(x_{i_{r-2}}+l_{i_{i-2}}^{*})
    \end{multline}
    If the next maximum variable is $x_{i_{r-2}}$ then we get 
    \begin{equation}
        F_1+F_2+F_3=(x_{i_{r-3}^{*}}+l_{i_{i-3}^{*}}^{*})(x_{i_{r-2}^{*}}+l_{i_{i-2}^{*}}^{*}).
    \end{equation}
    Extracting the coefficient of the maximum variable from the last equation (say $x_{i_{r-2}^{*}}$) we obtain
    $F_1=F_1^{*}$ since $x_{i_{r-2}^{*}}\in \ind(f_1/h)$ and thus can not belong to $F_2$ or $F_3.$ This ends the proof.\\
    If the maximum variable is not $x_{i_{r-2}}$ but for example $x_{i_{r-2}^{*}}$, the proof works exactly the same.
        \item If $x_{i_{r-3}}$ is belongs to $\ind(f_2^{*}/h^{*})$ we get 
    \begin{multline}\label{eq:Phffx3x3}
        (x_{i_{r-2}}+l_{i_{i-2}})\left(F_1+F_2+F_3\right)\\=
        (x_{i_{r-3}^{*}}+l_{i_{i-3}^{*}}^{*})(x_{i_{r-2}^{*}}+l_{i_{i-2}^{*}}^{*})(x_{i_{r-2}}+l_{i_{i-2}}^{*}+x_{i_{r-1}^{*}}+l_{i_{r-1}^{*}}^{*})
    \end{multline}
    We can apply the same arguments as in the previous case and obtain the wanted result.
         \end{enumerate}
    
    Now suppose our statement is true up to $\mu-1$ and let us show that it is also valid for $\mu.$ Using the same technique consider by absurd we have a common polynomial $P.$ 
    \begin{equation}\label{eq:Phh*mu}
        H(F_1+\dots F_{\mu})=H^{*}(F_1^{*}+\dots + F_{\mu}^{*})
    \end{equation}where the same notations as in the case of three variables is maintained. Suppose there are at least two maximum non-comparable monomials $f_1,f_2$ and $f_1^{*},f_2^{*}$, which implies $h=h^{*}$ and $f_1=f_1^{*},f_2=f_2^{*}.$ By sequentially extracting the coefficients of the maximum variables one by one we will show that we will have $F_i=F_{i}^{*}$ for some $i.$ Because there are many cases and the arguments are exactly the same each time (as in the case of 3 monomials) we shall demonstrate the concept for one case. Let $f_1=x_{i_1}x_{i_2},f_2=x_{i_3}x_{i_4}$ satisfying the previous conditions (maximum non-comparable pair of monomials). Then $i_3<i_1<i_2<i_4.$ Extracting the coefficient of $h$ and afterwards of $x_{i_4}$ from \eqref{eq:Phh*mu} gives 
    \begin{equation}
        x_{i_3}+l_{i_3}=x_{i_3}+l_{i_3}^{*}.
    \end{equation}
    Going to the next maximum variable, which is $x_{i_2}.$ Consider that $i_2\not \in\ind(f_3^{*}\dots f_{\mu}^{*}/h)$ and extracting this variable from the resulting equation gives
    \begin{equation}
        x_{i_1}+l_{i_1}=x_{i_1}+l_{i_1}^{*}.
    \end{equation}
    Continuing with $x_{i_1}$ gives $F_1=F_1^{*}$, and combined with the induction ends the proof. Switching the order of the indices in $f_1,f_2$ we get $F_2=F_2^{*}.$ \\
    The case of a single maximum monomial $f_1=f_1^{*}$ and $h\neq h^{*}$ can be treated as in the 3 monomials step. 
    % without loss of generality let $\forall i\in [1,l],\; f_i=f_i^{*}.$ Under our assumption $l\geq 1.$ For that, $P-\Alow_h\cdot h(\sum_{i=1}^l \Alow_{f_i}\cdot \frac{f_i}{h})$ is a common polynomial of the sets $\Alow_h\cdot h(\sum_{i=l+1}^{\mu} \Alow_{f_i}\cdot \frac{f_i}{h})$ and $\Alow_h^{*}\cdot hh^{*}(\sum_{i=l+1}^{\mu} \Alow_{f_ih^{*}}\cdot \frac{f_ih^{*}}{hh^{*}}).$ Using our induction this can be true only if $\forall i\in [l+1,\mu],\;f_i=f_i^{*}$, which contradicts our assumption and ends the proof. 
\end{IEEEproof}

Proposition \ref{pr:degree-two-minkowski-mu} combined with Proposition \ref{pr:no-colision-orbits-typeII} gives a closed formula for counting type II $\w_{\mu}$-weight codewords, which finalises the proof of Theorem \ref{thm:formula_typeII}.

%\section{Proof of Lemma \ref{lem:2dmin-orbits}}
\section{Proof of Theorem \ref{thm:count-all-subcodes-RM2}}
To demonstrate our result we need an intermediate result.
\begin{lemma}\label{lem:2dmin-orbits}
    Let $m$ be a positive integer and $l\in[0,(m-1)/2]$ such that there are $l$ monomials $f_i\in\I_2$ satisfying $\gcd(f_1,\dots f_l,x_j)$ for some variable $j.$ Then we have 
    \begin{multline}
        \label{eq:cardinal-orbit-2dmin}
        \left|\Alow\cdot x_j+\sum\limits_{i=1}^{l}\Alow\cdot f_i\right|\\=\left|\sum\limits_{i=1}^{l}\Alow\cdot f_i\right|\left|\Alow_{f_1\dots f_l}\cdot x_j\right|.
    \end{multline}
\end{lemma}
\begin{IEEEproof}
We will use induction and start with the first two cases.
\begin{itemize}
    \item $l=0\Leftrightarrow P=x_{i_1}$ and here the result is trivial.
    \item $l=1\Leftrightarrow P=(x_{i_1}+l_{i_1})(x_{i_2}+l_{i_2})+x_{i_3}+l_{i_3}.$ Suppose $l_{i_3}$ contains at least one variable from $x_{i_1}, x_{i_2}.$ We have
    \begin{align*}
        P&=(x_{i_1}+l_{i_1})(x_{i_2}+l_{i_2})+x_{i_3}+x_{i_2} + l_{i_3}^{*}\\
        &=(x_{i_1}+l_{i_1}+1)(x_{i_2}+l_{i_2})+x_{i_3}+l_{i_1} + l_{i_3}^{*}\\
        &=(x_{i_1}+l_{i_1}+1)(x_{i_2}+l_{i_2})+x_{i_3} + l_{i_3}^{\prime},
    \end{align*}
    where $l_{i_3}^{\prime}$ does not contain any of the variables $x_{i_1},x_{i_2}.$
    If $l_{i_3}$ contains both variables we have 
     \begin{align*}
        P&=(x_{i_1}+l_{i_1})(x_{i_2}+l_{i_2})+x_{i_3}+x_{i_2} + x_{i_1}+ l_{i_3}^{*}\\
        &=(x_{i_1}+l_{i_1}+1)(x_{i_2}+l_{i_2}+1)+x_{i_3}+l_{i_1} + l_{i_2}+1+ l_{i_3}^{*}\\
        &=(x_{i_1}+l_{i_1}+1)(x_{i_2}+l_{i_2})+x_{i_3} + l_{i_3}^{\prime},
    \end{align*}
     where $l_{i_3}^{\prime}$ does not contain any of the variables $x_{i_1},x_{i_2}.$
\end{itemize}

Suppose our statement is valid up to $l-1$ and let us demonstrate that it is also true for $l.$ Consider $P\in\Alow\cdot x_{i_{2l+1}}+\sum_{i=1}^{l}\Alow\cdot f_i.$ By definition we have $P=(x_{i_1}+l_{i_1})(x_{i_1}+l_{i_2})+\dots (x_{i_{2l-1}}+l_{i_{2l-1}})(x_{i_{2l}}+l_{i_{2l}})+x_{i_{2l+1}}+l_{i_{2l+2}}.$
We thus have 
\begin{multline*}
    P=(x_{i_1}+l_{i_1})(x_{i_1}+l_{i_2})+\dots (x_{i_{2l-1}}+l_{i_{2l-1}})(x_{i_{2l}}+l_{i_{2l}})\\
    +x_{i_{2l+1}}+\sum\limits_{\substack{j<i_{2l+1}\\j\not\in \{i_1,\dots, i_{2l}\}}}b_{i_{2l+1},j}x_j+\sum\limits_{\substack{j<i_{2l+1}\\j\in \{i_1,i_{2}\}}}b_{i_{2l+1},j}x_j+\varepsilon_{i_{2l+1}}\\
    =(x_{i_1}+l_{i_1}^{*})(x_{i_1}+l_{i_2}^{*})+\dots (x_{i_{2l-1}}+l_{i_{2l-1}})(x_{i_{2l}}+l_{i_{2l}})\\
    +x_{i_{2l+1}}+\sum\limits_{\substack{j<i_{2l+1}\\j\not\in \{i_1,\dots, i_{2l}\}}}b_{i_{2l+1},j}x_j\\+b_{i_{2l+1},i_1}l_{i_2}+b_{i_{2l+1},i_2}l_{i_1}+\varepsilon_{i_{2l+1}}^{*}+b_{i_{2l+1},i_1}b_{i_{2l+1},i_2}.
\end{multline*}

Using our induction hypothesis we can write $P-(x_{i_1}+l_{i_1}^{*})(x_{i_1}+l_{i_2}^{*})$ such that $l_{i_{2l+1}}^{\prime}$ does not contain any of the existing variables, and hence conclude our proof.
\end{IEEEproof}

Now, Lemma \ref{lem:2dmin-orbits} combined with Proposition \ref{pr:no-colision-orbits-typeII} and Theorem \ref{thm:formula_typeII} gives the wanted result.

\end{document}